\documentclass{emulateapj}
\slugcomment{{\sc Accepted to ApJ:} May 11, 2012}
\usepackage{graphicx}

\newcommand{\apg}{^{>}_{\sim}}

\newcommand{\farcc}{\mbox{\ensuremath{,\!\!^{\prime\prime}}}}

\shorttitle{Spatially-resolved Spectroscopy at $z$=0.7895}
\shortauthors{Frye et al.}

\begin{document}

\title{
Spatially-resolved HST Grism Spectroscopy of a Lensed Emission Line Galaxy at $z$$\sim$1$^1$}

\author{Brenda L. Frye\altaffilmark{2,3,4}, Mairead Hurley\altaffilmark{5}, 
David V. Bowen\altaffilmark{6}, Gerhardt Meurer\altaffilmark{7},   
Keren Sharon\altaffilmark{8}, Amber Straughn\altaffilmark{9}, 
Dan Coe\altaffilmark{10}, 
Tom Broadhurst\altaffilmark{11} and Puragra Guhathakurta\altaffilmark{12}}

\altaffiltext{1} {Based, in  part, on data obtained with the W. M. Keck Observatory, 
which is operated as a scientific partnership among the California Institute of 
Technology, the University of California, and NASA, and was made possible by
the generous financial support of the W. M. Keck Foundation.}
\altaffiltext{2}{Steward Observatory, Department of Astronomy, University of Arizona, 933 N. Cherry Avenue, Tucson, AZ 85721}
\altaffiltext{3}{Department of Physics and Astronomy, SUNY Stony Brook, Stony Brook, NY  11794-3800}
\altaffiltext{4}{Department of Physics \& Astronomy, University of San Francisco, 2130 Fulton Street, San Francisco, CA  94117}
\altaffiltext{5}{School of Physical Sciences, Dublin City University, Glasnevin 9, Dublin, Ireland; mairead.hurley5@mail.dcu.ie}
 \altaffiltext{6}{Department of Astrophysical Sciences, Peyton Hall, Princeton University, Princeton, NJ  08540}
\altaffiltext{7}{International Centre for Radio Astronomy Research, The University of Western Australia M468, 35 StirlingHighway, Crawley, WA 6009, Australia}
 \altaffiltext{8}{Department of Astronomy and Astrophysics,The University of Chicago, Chicago, Illinois 60637}
 \altaffiltext{9}{Astrophysics Science Division, Goddard Space Flight Center, Code 665, Greenbelt, MD  20771}
\altaffiltext{10}{Space Telescope Science Institute, 3700 San Martin Drive, Baltimore, MD  21218}
\altaffiltext{11}{Ikerbasque, Basque Foundation for Science, 48011 Bilbao, Spain}
\altaffiltext{12}{UCO/Lick  Observatory, Department of Astronomy and Astrophysics, University of California, Santa Cruz, CA  95064}

\begin{abstract}

We take advantage of gravitational lensing amplification by Abell 1689 
($z$=0.187) to undertake the first space-based census of emission line 
galaxies (ELGs) in the field of a massive lensing cluster.  Forty-three ELGs 
are identified to a flux of  $i_{775}$=27.3 via slitless grism spectroscopy.
One ELG (at $z$=0.7895) is very bright owing to lensing magnification by a 
factor of $\approx$4.5.  Several Balmer emission lines detected from 
ground-based follow-up spectroscopy signal the onset of a major starburst 
for this low-mass galaxy ($M_*\approx2\times10^9$M$_{\odot}$) with a high 
specific star formation rate ($\approx$20 Gyr$^{-1}$).  From the blue emission 
lines we measure a gas-phase oxygen abundance consistent with solar 
(12+log(O/H)=8.8 $\pm0.2$).  We break the continuous line-emitting region 
of this giant arc into seven $\sim$1kpc bins (intrinsic size) and measure 
a variety of metallicity dependent line ratios.  A weak trend of increasing metal
fraction is seen toward the dynamical center of the galaxy.  Interestingly, 
the metal line ratios in a region offset from the center by  $\sim$1kpc have a 
placement on the blue HII region excitation diagram with $f$([OIII])/$f$(H$\beta$) 
and $f$([NeIII])/$f$(H$\beta$) that can be fit by an AGN.
This asymmetrical AGN-like behavior is
interpreted as a product of shocks 
in the direction of the galaxy's extended tail, possibly instigated
by a recent galaxy interaction.

\end{abstract}


\section{Introduction}

Star formation in galaxies peaks at redshifts $z\sim2$ 
\citep{Reddy:05, Conselice:11}, with an overall decline in the global 
star formation rate density towards later cosmic times \citep{Lilly:96}.
Galaxies during the critical redshift range of 1$<$$z$$<$3 are actively 
converting gas into stars and at least
to some extent, building up central supermassive black holes
 \citep{Reddy:08, Somerville:09}.  Morphologically, these galaxies are 
 already well underway with assuming the familiar shapes of the Hubble 
 sequence \citep{Kriek:09}.   Given the unfortunate placement of major 
 star formation features split between the optical and infrared passbands, 
 building up a database of ELGs during this important cosmic epoch is slow.  

Sample sizes of ELGs at these intermediate redshifts are small and comprise
necessarily the brightest examples. Properties of these ELGs are determined 
typically from ratios of strong rest-frame optical emission lines. The values for 
the line ratios are persistently high compared to the standard excitation 
sequence for HII regions \citep{Erb:06a, Straughn:09, PerezMontero:09,
Xia:11, Trump:11}.  In turn the elevated line ratios are generally accompanied by higher 
ionization parameters and electron densities, indicating that physical conditions 
may be different from the local universe \citep{Liu:08, Brinchmann:08,Lehnert:09,
Richard:11}.  Spatially-resolved observations can assist with the search for an
explanation of these consistently high nebular line ratios.  In one case of a 
field galaxy at $z=1.6$ HDF-BMZ1299, integral field unit observations have 
revealed line ratios in the innermost $\sim$1.5 kpc that are best-fit by an AGN.  
This measurement would not have been achievable  in the spectrum integrated over the 
whole object \citep{Wright:10}.  Similarly, a Hubble Space Telescope (HST) 
Wide Field Camera 3 (WFC3)
grism study of ELGs from CANDELS \citep{Grogin:11, Koekemoer:11} data in
GOODS-S reveal the likelihood of weak AGN activity in $z$$\sim$2 low mass, 
low metallicity galaxies \citep{Trump:11}.

It is  feasible to use low resolution grism spectroscopy with HST ACS  to study 
ELGs and also galaxies with strong breaks (bulges, 
Lyman-break galaxies).  This approach has been used extensively by the 
Grism Advanced Camera Program for Extragalactic Science (GRAPES) and 
Probing Evolution and Reionization Spectroscopically (PEARS) Treasury 
teams (PI:  Malhotra).  For example, \citet{Hathi:09} studied the stellar populations
of late-type galaxies at $z$$\sim$1 in the Hubble Ultra Deep Field (HUDF) using
low resolution grism spectroscopy.   They identified the bulges in a sample of 
34 galaxies by a combination of their prominent 4000 \AA \ break and visual
morphologies. They measured stellar ages in the bulges that are similar to 
stellar ages in the inner disks and used this information to constrain galaxy 
formation mechanisms.  \citet{Ferreras:09} measured galaxy properties 
for a sample of 228 galaxies at $z$$\sim$1 selected by morphology and HST 
ACS grism spectroscopy to be early-types.  They modeled the star formation 
histories and found the galaxy formation epoch to correlate strongly with stellar 
mass in massive early-type galaxies.  \citet{Xia:11} reported on a comparison of 76 
ELGs in Chandra Deep Field South (CDFS) acquired with both the ACS grism 
and ground-based spectrographs covering similar wavelengths.  They compared 
the grism 
redshift estimates for a typical case of a single emission line plus broad-band
photometry with the higher resolution ground-based spectroscopy and 
successfully recovered the original grism redshift estimates.
In yet a different use of the ACS grism, \citet{Nilsson:11}
undertook a study of Lyman-break galaxies (LBGs) at $z$$\sim$1 detected in the
UV by GALEX.  They used grism spectra covering the 4000 \AA \ break
to measure more accurate positions of this break than could be achieved
by the
broad-band colors alone, which they then used to inform the SED models.  The
measurements yielded physical characteristics of their low-redshift 
sample of LBGs that were
similar to LBG properties at higher redshifts.

More recently, grism spectroscopy using WFC3 on HST Early Release 
Science program \citep{Windhorst:11} has yielded the discovery of  
48 ELGs to a limiting magnitude of $m$=25.5 (AB).  They acquired one  
field observed with the G102 and G140 grisms, each at a 2-orbit depth 
\citep{Straughn:11}.  Specific star formation rates (sSFR) 
were measured based on the SED fits that were low for the 
highest mass galaxies and that evolved with redshift, in general agreement
with galaxy downsizing.
\citep{Feulner:05,Bauer:05,Elbaz:07, Rodighiero:10, Damen:09, Zheng:07}.

The introduction of gravitational lensing to grism analyses 
is  useful because it boosts the brightnesses and 
sizes of all objects in the background, allowing for the study of individual 
sources of line emission at higher signal-to-noise and higher spatial 
resolution.   For the case of an emission line on top of stellar continuum
in particular, the effect
of field dilution is to smear out the continuum flux over more pixels.  
At the same time the compact star forming regions or galactic nuclei are also
magnified but effectively remain unresolved.  Thus the extended continuum
is diluted with respect to the emission lines, and the detection threshold is
lowered to include weaker emission line systems.
Assisted by the lensing effect, the first metallicities at intermediate redshift 
are being measured directly \citep{Yuan:09, Rigby:11}.  Also spatially-resolved 
spectroscopy is achieved, enabling measurements of star formation
properties across the disk \citep{Jones:10b, Hainline:09, Frye:07}.

The choice of a large, cluster-sized lens offers magnification of all galaxies in 
the background over fields of $\sim$2-3 arcmin in radius.  Abell 1689 ($z=0.187$,
measured by Frye et al. 2007) with its large tangential critical curve of 
50$^{\prime\prime}$ is one of the most massive and well-studied clusters 
\citep{Broadhurst:05, Limousin:07, Coe:10}.  The galaxies situated behind 
massive clusters have notoriously low surface brightnesses, and as such a 
space-based platform like HST is necessary for undertaking detailed and 
spatially-resolved studies of ELGs with fields of view larger than can be achieved 
with integral field unit spectroscopy from the ground.  HST has already proved 
highly successful at identifying ELGs in a survey mode at intermediate redshifts
\citep{Xu:07,Straughn:08, Straughn:09, Straughn:11, Trump:11}, and at high 
redshifts \citep{Malhotra:05,Rhoads:09}.  In addition to the wide field of view,
spatially-resolved spectroscopy is achieved with a resolution equal to our
minimum aperture extraction width five rows or 0.25${\farcc}$
and is acquired in the absence of competing strong atmospheric skylines.

Here we present a new census of ELGs in the field of the lensing cluster A1689
with the G800L grism on Advanced Camera for Surveys (ACS) on HST.
This is the first space-based slitless emission line survey centered on a 
massive lensing cluster.
One new ELG at $z\approx 0.79$ is a star forming galaxy which we designate 
by the paper reference Frye et al.~2012 object 1, which shortens to `F12$\_$ELG1.' 
This giant arc with $\sim8^{\prime\prime}$ extent is rare for showing 
continuous line emitting region in several features over half its optical extent.  
We have taken both space- and ground-based spatially-resolved spectroscopy 
of this one bright ELG with a total magnitude integrated over the galaxy 
image of $i_{775} = 20.56 \pm 0.01$.  This object is suitable for exploring 
the variations in physical properties across the galaxy disk.

The paper is organized as follows. In \S2 we present our imaging and spectroscopic 
sample and data reduction techniques. In \S3 we give results relating to ELGs in
the cluster.  In \S4 we report results for new giant arc F12$\_$ELG1.    In \S5 we 
use a lens model to construct a 1d magnification profile, compute individual
magnifications for galaxies of interest, and to generate a source plane image for 
F12$\_$ELG1.  In \S6 we discuss other field galaxies of interest, including
 another new sample ELG with closely spaced sources of line emission 
 which we call `F12$\_$ELG2.'  The conclusions appear in \S7.  We assume a cosmology 
with $H_0=70$ km s$^{-1}$ Mpc$^{-1}$, $\Omega_{m,0} = 0.3$, $\Omega_{\Lambda,0} = 0.7$.


\section{Observations \& Reductions}

\subsection{ACS Grism Data}

The central portion of A1689 is observed with the G800L grism with ACS on HST
along with 
the broad band (F475W, F625W, F775W, and F850LP) exposures presented in
\citet{Broadhurst:05}.
A single pointing is used at one position angle in a 7.1 ks exposure (3 orbits).
The resulting dispersed image covers a wavelength range of 5700 - 9800 \AA \  
at a resolution of $R \sim 90$. 
The analysis of these images is performed using software discussed in 
 detail by \citet[hereafter M07]{Meurer:07}.
 M07 present G800L grism observation of a well-studied unlensed field, the Hubble 
Deep Field North (HDF-N), to a similar depth as our observations.  They reduce and analyze this dataset
in two different ways:  Method A: aXe selection and reduction \citep{Kummel:09}, which is similar to the GRAPES team pipeline, 
and Method B: `blind' emission line source detection.  Method A yields a 
similar set of sources as Method B, with the
latter additionally yielding $>$50\% more sources and more cases of multiple
emission line sources ({\it i.e.}~star forming regions) per object. 
A salient feature of Method B is its emphasis on finding emission lines
in the 2d dispersed frames.  
This approach enables lone emission lines with weak
galaxy continuum flux to be detected 
against the high background that is characteristic of grism images.   For our
aims to do spatially-resolved spectroscopy and to maximize the number of
emission line sources, we perform the reductions using Method B. 
The primary steps of the reduction algorithm are discussed below.
 
 The initial data reduction is performed with the Space
Telescope Science Institute (STScI) ACS CCD reduction pipeline 
CALACS \citep{Hack:99}, from which we use the cosmic ray rejected ``CRJ'' 
G800L images along with the individual flatfielded ``FLT'' broad band images
 (F475W, F625W, F775W, and F850LP).  Since CALACS does not flatfield G800L exposures 
we apply our own corrections for pixel-to-pixel variations using a standard F814W flat 
from the STScI archive.  To remove small dithers between exposures and
to correct for geometric distortion we employ the ACS team pipeline Apsis 
 \citep{Blakeslee:03}.  These initial reduction steps produce geometrically-corrected 
 and aligned grism images, which are processed further as described below.
 Similarly, Apsis is used to align and process the set of broadband images;
 the weighted sum of these is what we designate as the ``direct" image.  
  
\begin{figure}
\includegraphics[viewport=-20 -6 200 465,scale=.62]{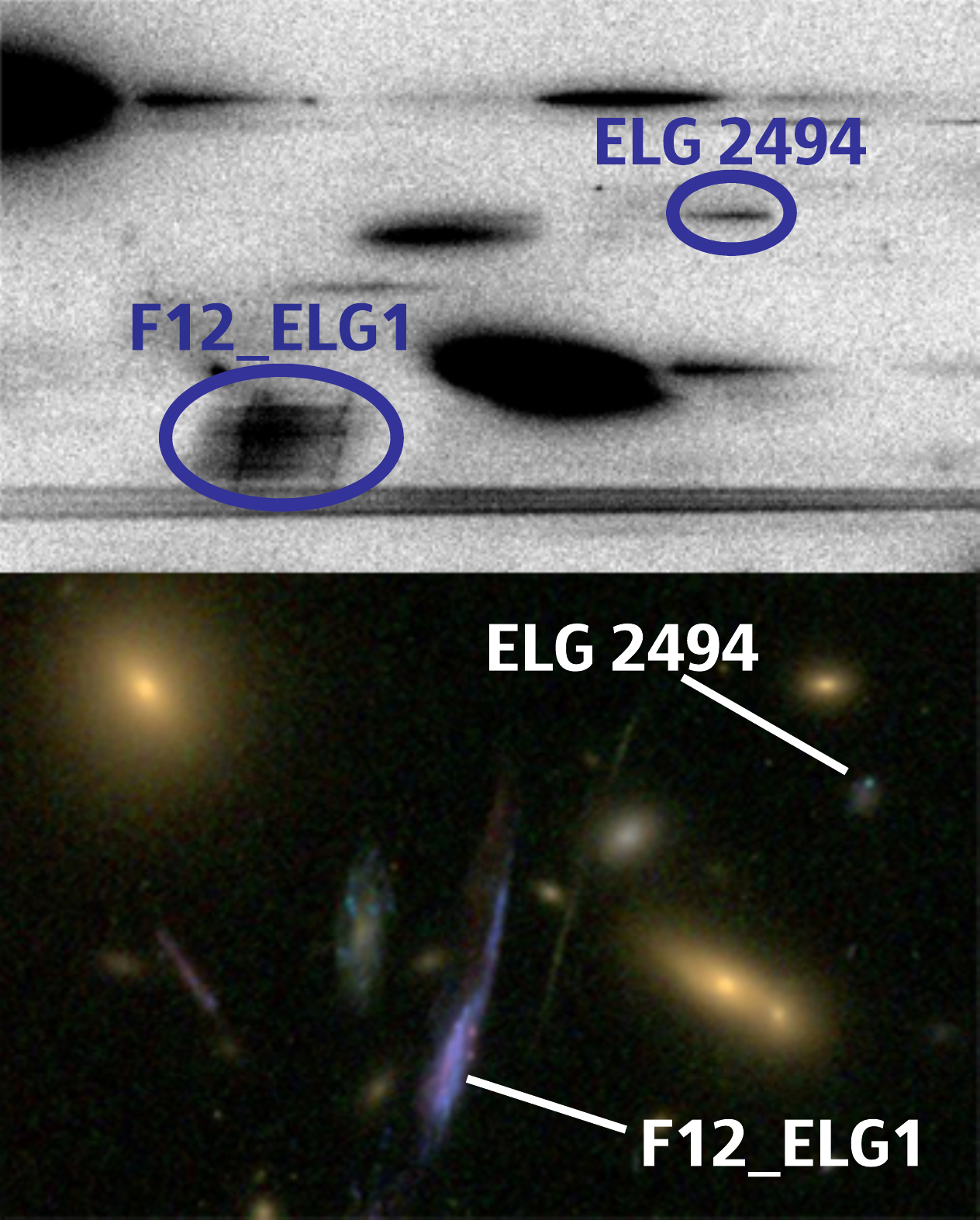}
\caption{Upper panel: dispersed image of a portion of the central region of A1689,
including the giant arc F12$\_$ELG1 that is a primary result of this paper and a 
cluster member, ELG 2494.  The dispersion direction is horizontal and the spatial
direction is vertical.  Note the spatially-extended emission in F12$\_$ELG1.
Three ELSs are found in this single ELG 
and their 1d spectra are shown in Figures \ref{figbigarc}, \ref{family}, and \ref{oii}.
Lower panel:  color 
image of the same portion of A1689 taken with the HST ACS $gri$ filters.
 \label{2dspec}}
\end{figure}

\begin{figure}
\includegraphics[viewport=0 0 0 580,scale=0.5]{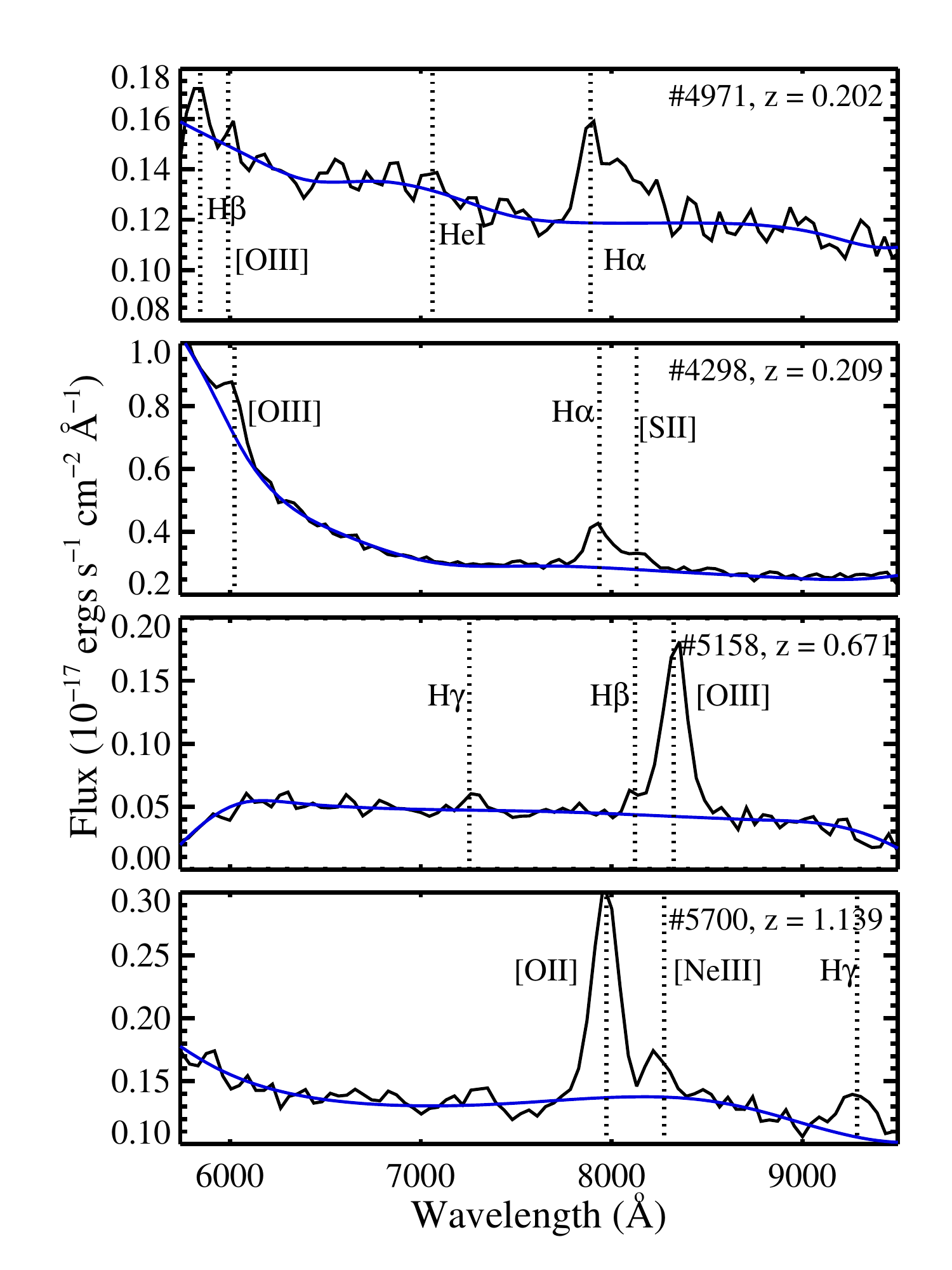}
\caption{Sample spectra from our G800L ACS grism spectral catalog.  
The spectrum for ELG 4971 shows a detection of H$\alpha$ in a cluster
member at $z$=0.202 (top panel).
Despite the low spectral resolution of the grism, redshifts can still be 
derived even from closely spaced nebular emission features.  For example, 
H$\alpha$  
be distinguished from [SII] in some cases (ELG 4298), H$\beta$ can be distinguished
from [OIII] in ELG 5158, 
and [NeIII] can be seen clearly with respect to 
[OII] in ELG 5700.  
Details concerning these results appear in Table~\ref{ID}.
 \label{samples}}
\end{figure}

Galaxies with strong emission lines (ELs) should be the easiest sources 
to identify in slitless images like our G800L data.  Method B of M07 requires
no prior knowledge of the location 
of the emitting source, although the direct image is needed to find the precise
position of the emission.  Moving forward from 
the initial reductions discussed above, we next subtract 
off a 13x3 median smoothed version of the dispersed image, revealing the 
emission line(s).  We also subtract off a smoothed version of the direct 
image from itself.  The result of this high-pass filter is to remove slowly-varying
background levels, including galaxy continuum, starlight from 
galaxy neighbors and other sources.  As detailed in M07 we have determined
a geometric distortion solution  relating positions in the direct image to the corresponding positions 
of the zeroth order in the grism image.  This solution along with the measured flux 
scaling between direct and zeroth order images is used to mask out all significant flux from
zeroth order images in the grism frame. SExtractor \citep{Bertin:96} is used to find the 
emission line sources in this masked and filtered grism frame.  A five row segment of both of the
the filtered dispersed and direct image are extracted, collapsed to 1D spectra and 
cross-correlated.  The peak in the cross correlation gives 
the location of the emitting source and an estimated line wavelength.  The aXe package is then used to extract a 1D
spectrum from the dispersed image (prior to any filtering) using the location we found as 
the adopted source position.   Gaussian fits to the lines in 
this spectrum are used to determine their final wavelength and flux values.
A portion 
of the dispersed image that includes the spatially-resolved line emission for 
ELG F12$\_$ELG1 at $z$=0.79 appears in Figure~\ref{2dspec}.  We 
present the 1D spectrum of F12$\_$ELG1 in \S4 and \S6.   
Figure~\ref{samples} shows 1D spectra for other representative ELGs in our sample.  Some line pairs can be distinguished which assist 
with line identifications.  Notably, nebular lines [OIII] and H$\beta$ are marginally resolved 
in the new galaxy ELG 5158, and the close line pair H$\alpha$ and [SII] can both be 
identified in the new galaxy ELG 4298.  
 
We refer to an emission feature in the dispersed image as an EL (typically
H$\alpha$, H$\beta$, [OIII] or [OII]).  
The corresponding  position in the  direct image is called an emission line 
source (ELS).  An ELS is typically an HII region or galaxy
nucleus, and in some cases there is more than one EL per ELS (e.g.\ [OII] and [OIII]).
The galaxy containing the ELS(s) is an emission line
galaxy (ELG).  If the galaxy is spatially resolved, there may be multiple
ELSs per ELG.
We identify a total of 43 galaxies (ELGs), 
52 ELSs and 66 ELs.  Somewhat surprisingly, three-fourths of the ELGs are new 
in this well-studied cluster field.  The spectroscopic results appear in 
Table~\ref{ID}.  We identify nine ELGs in the cluster, 30 ELGs in the 
background, and five ELGs in the foreground.
As the number of ELSs is larger than the number of ELGs, some ELGs
have multiple H II regions spatially-resolved by the grism.  In Table~\ref{multiples} 
we separate out the ELGs with two or more ELSs.  The details concerning
our line list and emission line properties sorted by species are given in the Appendix.   

 \subsection{Imaging and Additional Spectroscopy}
 
Extensive ancillary data exist to support the grism observations.  Images 
of the central portion of A1689 were taken in several bands, as follows:  
$U$ (DuPont Telescope, Las Campanas), $B$ (NOT, La Palma), 
$V$ (Keck II LRIS), $I$ (Keck II LRIS), 
$g_{475}$$r_{625}$$i_{775}$ and $z_{850}$ (HST ACS), and $P_{3.6}$ and 
$P_{4.5}$ (Spitzer IRAC).   Spectroscopy of hundreds of arclets have been 
acquired at several large ground-based observatories.  The existing photometry 
and spectroscopy are discussed in detail elsewhere 
\citep{Broadhurst:05, Frye:02,Frye:07, Frye:08}.

Additional spectroscopy was obtained for F12$\_$ELG1 at $z$=0.79 
in 2010 May with the DEep Imaging
Multi-Object Spectrograph (DEIMOS) on Keck II \citep{Faber:03}.  The observations
were made through a 1 arcsec slit width with the 1200 line mm$^{-1}$
grating blazed at 10.16$^{\circ}$ and set to a central wavelength of
7800 \AA.  A combination of 12 slitlets were placed together to
construct the longslit mask ``Long1.0B."  Two 300 s exposures were
taken during dusk twilight.  The seeing was 0.5-0.6$^{\prime\prime}$
FWHM.  Long1.0B provided a resolution of $R$=5870 measured from an
isolated night sky line at 7571.75 \AA.  Observations were also
acquired on the Magellan Telescope I (Walter Baade) in 2009 March and
2010 March on the Inamori Magellan Areal Camera and Spectrograph
(IMACS).  One multislit mask was used with 35 targets 
using the 600 lines mm$^{-1}$
grism at a blaze angle of 14.67 \AA \ and a central wavelength of
8410 \AA \ as a part of a different program (PIs Malhotra and Rhoads).  The grism provided a resolution of $R$ = 2300.
Eight 1800 s exposures were taken in a single position.

\section{The Cluster}

\begin{figure}
\includegraphics[viewport=30 0 200 420,scale=0.45]{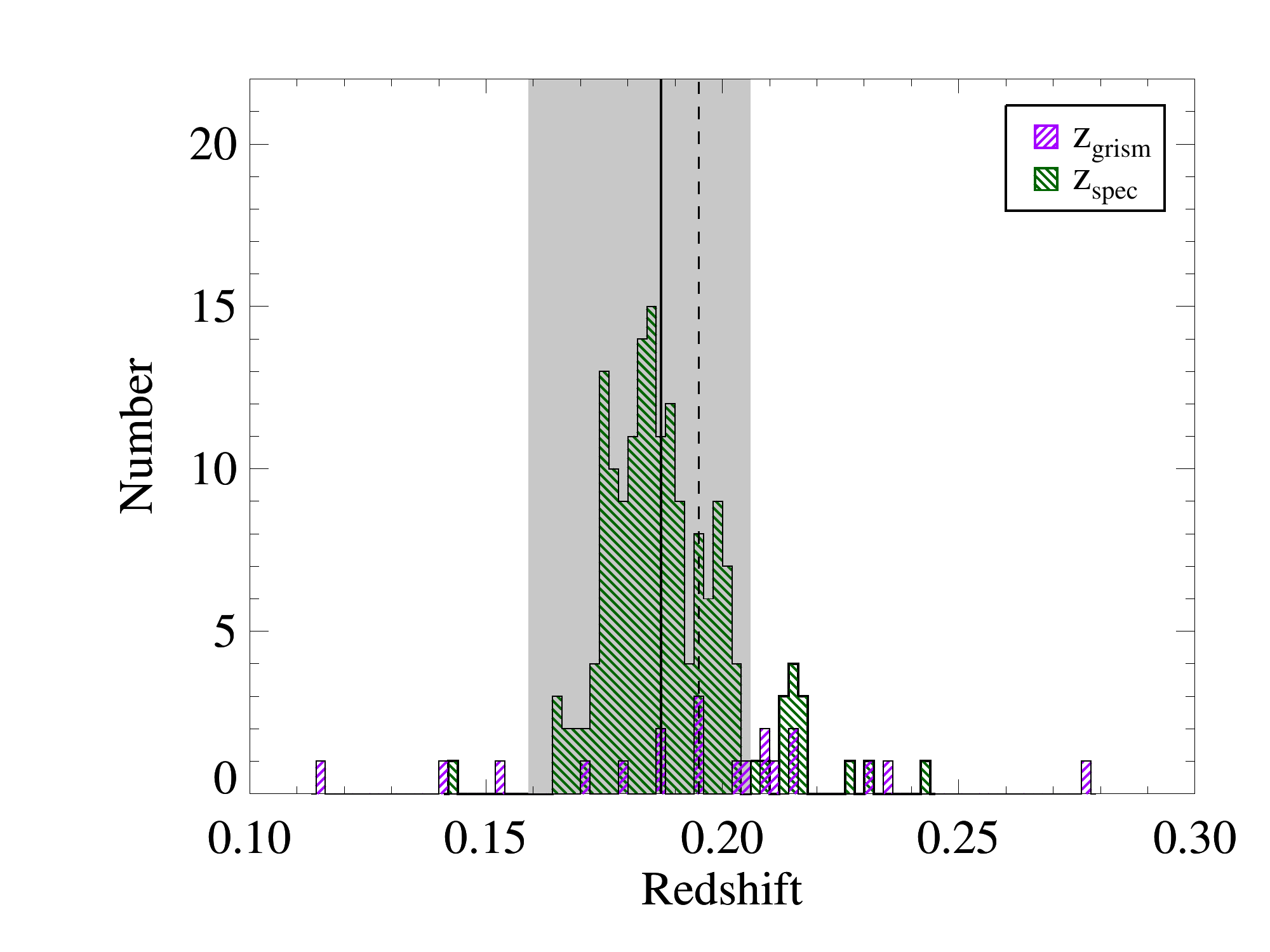}
\caption{Histogram of HAEs from this sample (purple positive-slope 
hashed region), compared to that of all published spectroscopic redshifts 
in A1689 (green negative-slope hashed region).  There is an extended 
tail towards higher redshifts  which may be indicate substructure and/or 
cluster infall.  The HAEs that satisfy the redshift criteria for cluster membership
are enclosed by the gray filled region. The mean redshift of all 
known cluster members is given by the solid line and the mean redshift 
of our sample cluster members is given by the dashed line.
\label{zhist}}
\end{figure}

\begin{figure}
\includegraphics[viewport= 10 110 300 675,scale=0.43]{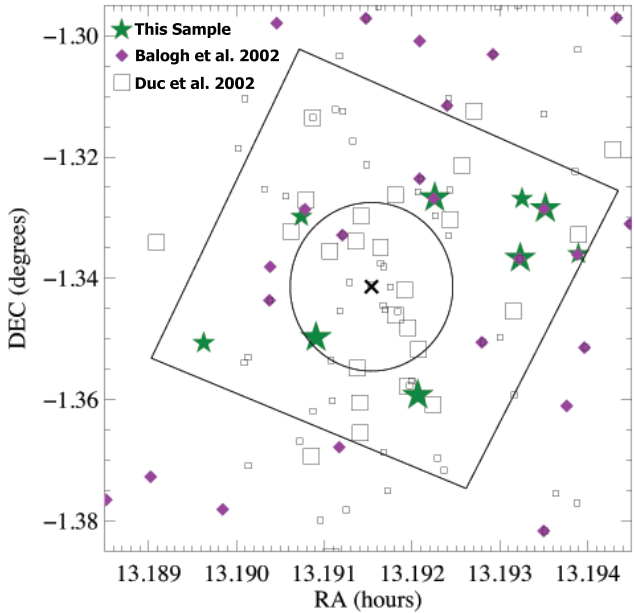}
\caption{Map of HAEs in A1689 from our census (green stars) and
from the sample of \citet{Balogh:02} (purple diamonds).  
Our HAEs are further organized by flux, with small and 
large symbols representing objects with $log(f_{H\alpha})$$<$$-16.0$ and 
$log(f_{H\alpha})$$\geq$$-16.0$ respectively. The HAEs are well-distributed
azimuthally, and appear to be underpopulated at smaller cluster-centric radii. The
positions of cluster members with absorption redshifts from \citet{Duc:02}, 
and the few that could be found in the literature, are indicated by the open-squares
and concentric-squares, respectively.  The canonical Einstein ring is indicated by the large
 black circle, with cluster center at `X.' 
 \label{scatter}}
\end{figure}

We identify 8 ELSs that are H$\alpha$ emitters (HAEs)  in the redshift range
0.159$<$$z$$<$0.206, which is the 3$\sigma$ velocity dispersion
redshift criteria  established by \citet{Balogh:02}.   There are three H$\alpha$
emitting sources in the foreground of A1689, and ten H$\alpha$ emitting 
sources with $z$$>$0.206.  There is a high velocity tail in the redshift range 
0.207$<$$z$$<$0.215 that includes five ELSs that are HAEs   (Fig.~\ref{zhist}).  
Our sample is distinguished from the larger ground-based H$\alpha$ galaxy 
survey of \citet{Balogh:02} primarily by our fainter flux limit of $i_{AB}=27.3$. 
Six of the galaxies that are H$\alpha$ emitters with redshifts 
0.159$<$$z$$<$0.215 are new to the literature:  ELG 6621a (serendipitous 
discovery, see Appendix), ELG 11226, ELG 4298/4251 (ELS 4298 and ELS 4251),
ELG 2494, ELG 10412, ELG 1507.

Figure~\ref{scatter} shows the locations of our sample cluster members 
(green stars) and of H$\alpha$ emitters drawn from the literature 
(purple diamonds).  The [OII], [OIII] and Balmer line emitters from the large 
sample of \citet{Duc:02} appear as large open squares.  The positions of 
other cluster members are also marked (small open squares), drawn largely
from the literature as compiled in \citet[their Appendix]{Frye:07}.  The large 
tilted rectangle marks the ACS field of view and the black circle shows the 
position of the canonical 50 arcsec Einstein ring for A1689.  The HAEs 
presented in this paper have measured line fluxes of $-14.82 <$ log $f < -16.61$.
The symbols are organized by total flux, with small and large symbols identifying
$log(f_{H\alpha}) \geq -16.0$.  Using the relation in \citet{Kennicutt:98} we 
compute star formation rates of $0.017 < $SFR $< 1.1 $M$_{\odot}$ yr$^{-1}$.

The positions of the ELGs cover the far-field reasonably well.  Of the 30 
H$\alpha$ emitters comprising this sample plus those drawn from the literature,  
only one is located close in to the cluster center.  This is a new ELG presented 
in this paper, ELG 4298/4251, which with $i_{775} = 21.74 \pm 0.01$ is not 
reported in the large H$\alpha$ survey of \citet{Balogh:02} as it is fainter than 
their $I$-band magnitude-limit of 19.3.  The preferential placement of H$\alpha$
emitters in the cluster outskirts appears to satisfy the general trend of increasing 
star fraction as density decreases, and is biased as a result of confusion
from a high source density in the cluster interior and small number statistics.  
It is interesting to note that there is an excess of galaxies detected at 100 $\mu$m 
with Herschel  that is distributed as a swath running NE-SW across 
their  field 
\citep{Haines:10}.  Evidence of filamentary structure is not seen in our galaxy
census which is less than one-tenth the field size of their sample.  

\section{F12$\_$ELG1:  The Giant Arc at $z$=0.7895}

\subsection{Galaxy Properties Measured from the HST Grism}

\begin{figure}
\includegraphics[viewport=  -20 -130 200 100,scale=0.17]{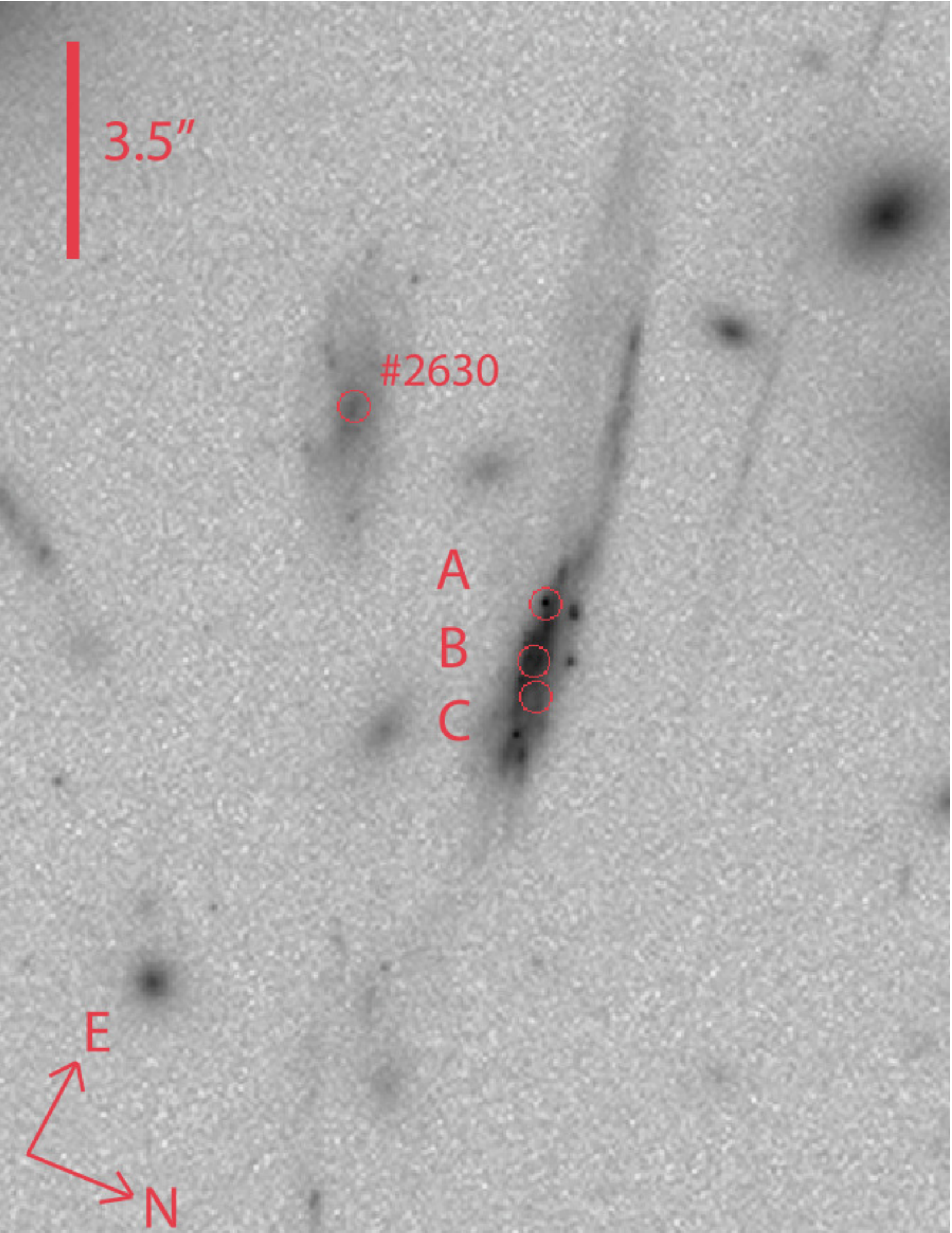}
\includegraphics[viewport=  -160 -50 200 430,scale=0.28]{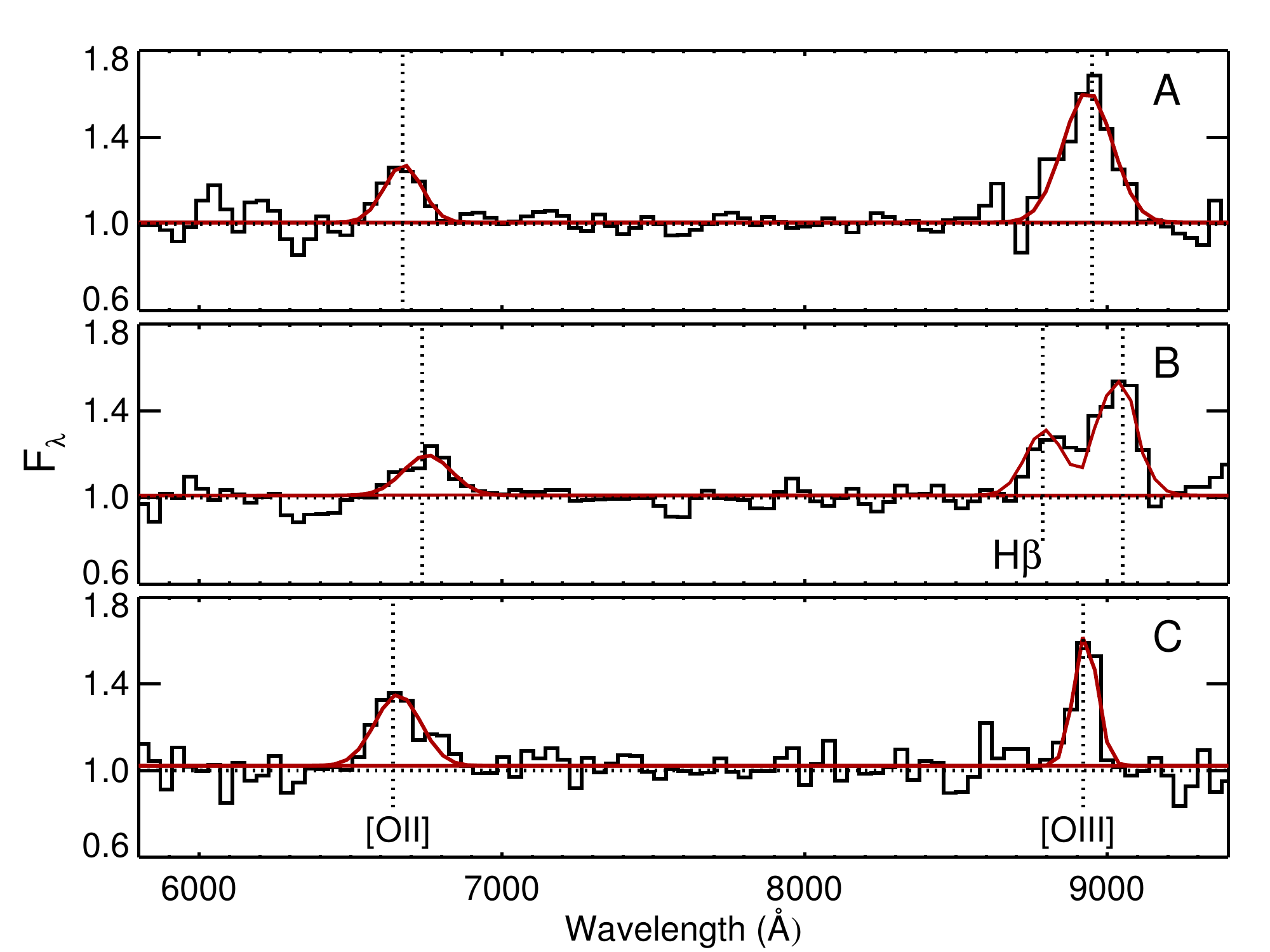}
\caption{$i_{775}$ image and flux-normalized G800L spectra are shown for
the bright giant arc F12$\_$ELG1 at $z$=0.790.  The spatially-resolved
spectroscopy corresponds to the three sources of line emission along the 
long axis of this galaxy, ELS 20002 (`A'), ELS 10638 (`B'), and ELS 10640 (`C').
Components B and C are  0.27$\arcsec$ apart, corresponding to an intrinsic
separation of only 0.5 kpc.  The flux of [OIII] relative to H$\beta$ is highest in 
the center, and from follow-up spectroscopy we find a region near the center to be 
best-fit by an AGN (See \S6.1 and Fig. ~\ref{trends}).  The shift in the line 
peaks between components A, B, and C is a result of large uncertainties in 
the zero-point of the wavelength solution. We detect emission in a second galaxy near in projection to F12$\_$ELG1, labeled as ELS 2630 at $z=0.335$.  
\label{figbigarc}}
\end{figure}

The image and spatially-resolved G800L spectra of F12$\_$ELG1 with
a visual extent of $>8$ arcsec and a magnification provided by the
cluster of a factor of 4.5 appear in Figure~\ref{figbigarc}.  This is
one of a handful of the brightest star forming galaxies at $z\sim1$
(M$_{B}$=$-21.3$), allowing for a study of one ELG in detail at
intermediate redshift. The HST grism spectroscopy shows three separate
emission line sources (ELSs) as ELS 20002 (`A'), ELS 10638 (`B'), and
ELS 10640 (`C').
Each ELS has at least two emission lines: [OII]
$\lambda\lambda$3726, 3729 and [OIII] $\lambda\lambda$4959, 5007, with
the central knot `B' also showing H$\beta$ line emission.  The line
fluxes and rest equivalent widths are given in Table~\ref{ID}.  A fit
to the line centroids of the emission features in Component A yields a
redshift of $z$=0.790, confirming the redshift in \citet{Duc:02}.
Knots A and B have a separation at the source of 1.5 kpc, while knots
B and C are separated by only 0.5 kpc.  There appear to be significant
redshift differences between knots A, B, and C.  These are likely to
be owing to misidentification of the precise $x$ position of the ELSs,
which will translate into a wavelength error.
 
\subsection{Galaxy Properties Measured from Ground-based Spectroscopy}

In our Keck and Magellan spectra of F12$\_$ELG1 we see line emission from
star forming regions across a continuous 4${\farcs}$ From the 1D
spectroscopy of this extended line emission we recover all the
emission features of the HST grism spectrum, and detect as well
additional emission lines.  We identify: [O II]$\lambda$3726, [O
  II]$\lambda$3729, [O III]$\lambda$4959, [O III]$\lambda$5007,
H$\beta$ and H$\gamma$, with H$\delta$, H$\epsilon$, H8, and [Ne
  III]$\lambda$3869, and weak [C III]/[C IV]$\lambda$4650 (in the Magellan
spectrum).  While the results are similar for our two ground-based
datasets,  we will focus primarily
on the Keck spectrum herein
with its higher spectral resolution and flux calibration taken with a
standard star at the time of the observations,.  The line fluxes
for the strong emission lines for the Keck spectroscopy appear in
Table~\ref{fluxes}.

\begin{figure}
\includegraphics[viewport = 0 0 200 485,scale=0.6]{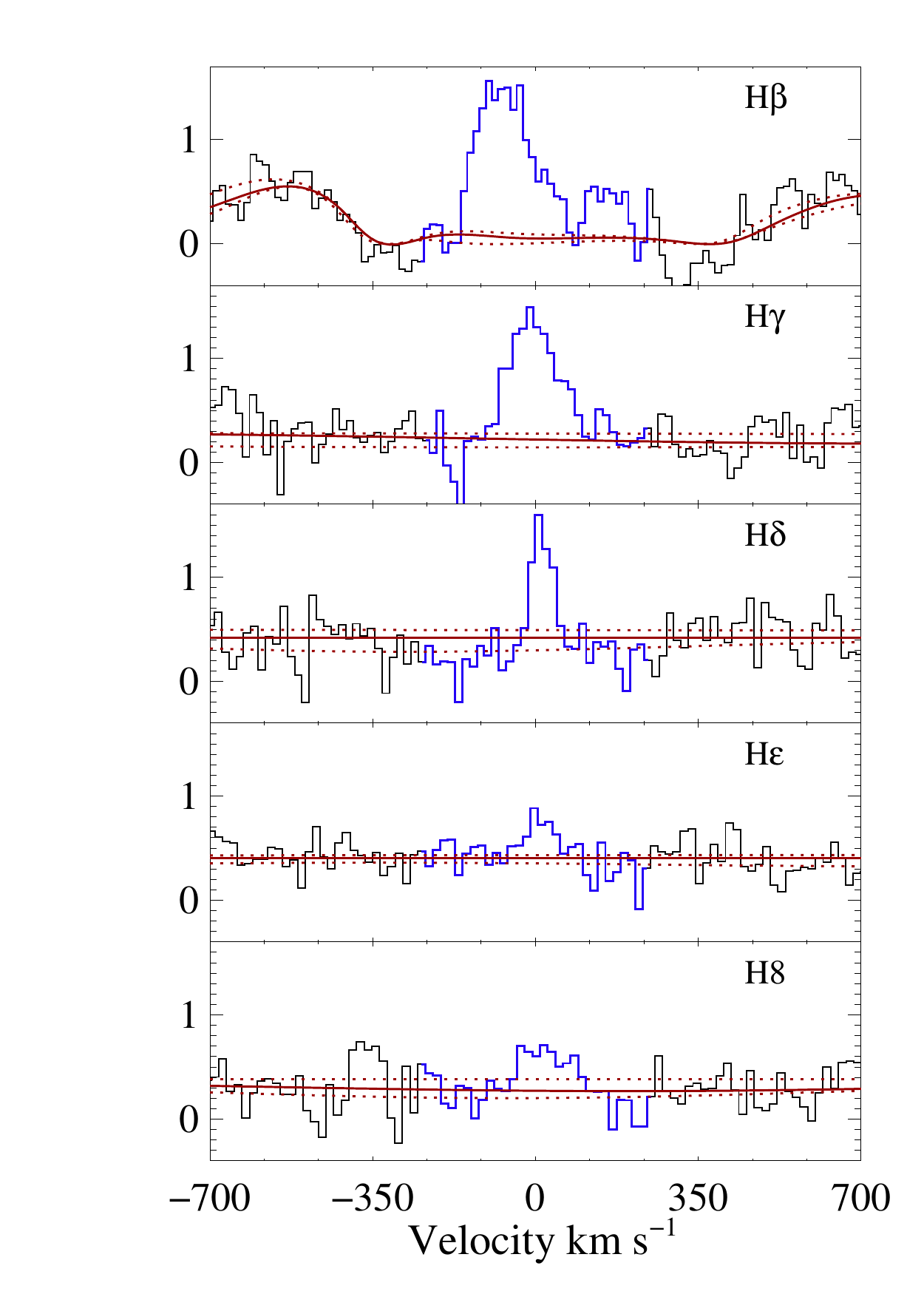}
\caption{Stackplot of the Balmer family of emission lines for F12$\_$ELG1, as
  labelled.  Most notable is the rare detection of Balmer transmission
  lines continuously in emission from H$\beta$ through H8, indicating
  a young starburst event.  The continuum fit plus uncertainties are
  given by the red solid plus red dashed lines.  The velocity range of
  the line (in blue) corresponds to $-250$ km s$^{-1}$$<$$v$$<$$+250$
  km s$^{-1}$ with reference to the systemic redshift of $z=0.7895$.
  These data are from our Magellan follow-up spectroscopy and the units
  are in normalized counts.  All spectra have been corrected for
  extinction by dust.  \label{family}}
\end{figure}

A Gaussian fit to [OII] and [OIII] in our Keck spectrum yields a new
systemic redshift of $z$=0.7895.  The line width is estimated from the
fit to the [OIII] $\lambda$5007 line, and is found to be $\Delta v =
500$ km s$^{-1}$ after subtraction of the instrumental resolution.
This is the line width set for all other emission features.  To
account for the slightly asymmetric line profile of the [OIII]
$\lambda$5007 line owing to an adjacent sky line, the flux value is
taken from the Gaussian fit rather than from the data values.  The
[OIII] $\lambda4959$ line also suffers from its unfortunate placement
relative to a skyline.  In this case the well defined ratio of [OIII]
$\lambda$5007 to [OIII] $\lambda4959$ of 3:1 is used for any
calculations involving the sum of the fluxes or equivalent widths of
these two lines.  For all other line features, the values are measured
directly from the data.  The Balmer transition lines H$\beta$ through H8
are shown in Figure~\ref{family}.  The velocity range over our 500 km
s$^{-1}$ measurement width is shown in blue and varies from $-250$ km
s$^{-1}<v<+250$ km s$^{-1}$ with reference to the systemic redshift of
$z=$0.7895.  This large family of Balmer features all in emission is
rare for an extragalactic source and signals the early stages of a
major starburst.

Our Keck spectrum has at best only a slight continuum break at
rest-frame 4000\AA. \citet{Balogh:99} define this break
$D_n(4000)$ as the ratio of the average flux density in the wavelength
band 4000-4100\AA\ to the 3850-3950 \AA\ band.  We measure a break
index of $D_n(4000) = 0.96 - 1.0$ after first masking out the strong
[Ne III]$\lambda$3869 and H8 emission lines. This near lack of a
continuum depression indicates a dominant population of hot young
stars. The population synthesis models of \citet[hereafter
  BC03]{Bruzual:03} provide a value of $D_n(4000)$ for each spectral
energy distribution (SED; see \S4.3 for a description of our model).
For a reasonable subset of models over four metallicities (0.2$Z_\odot$,
0.4$Z_{\odot}$, $Z_{\odot}$, and 2.5$Z_{\odot}$) and a range of star
formation histories, we have set an initial constraint on the age of
the dominant stellar population of the galaxy to a range of 6 Myrs $
\leqslant t \leqslant 100$ Myrs. Note that any contribution from AGN
continuum light would operate to raise the value for $D_n(4000)$, and
hence lead to an underestimate in the age of the galaxy
\citep{Kauffmann:03b}.  We infer from spatially-resolved analysis 
that this object is most likely dominated by star
formation but also shows evidence of a harder ionizing source (see \S6).

 \begin{figure}
\includegraphics[viewport = 0 0 200 360,scale=0.65]{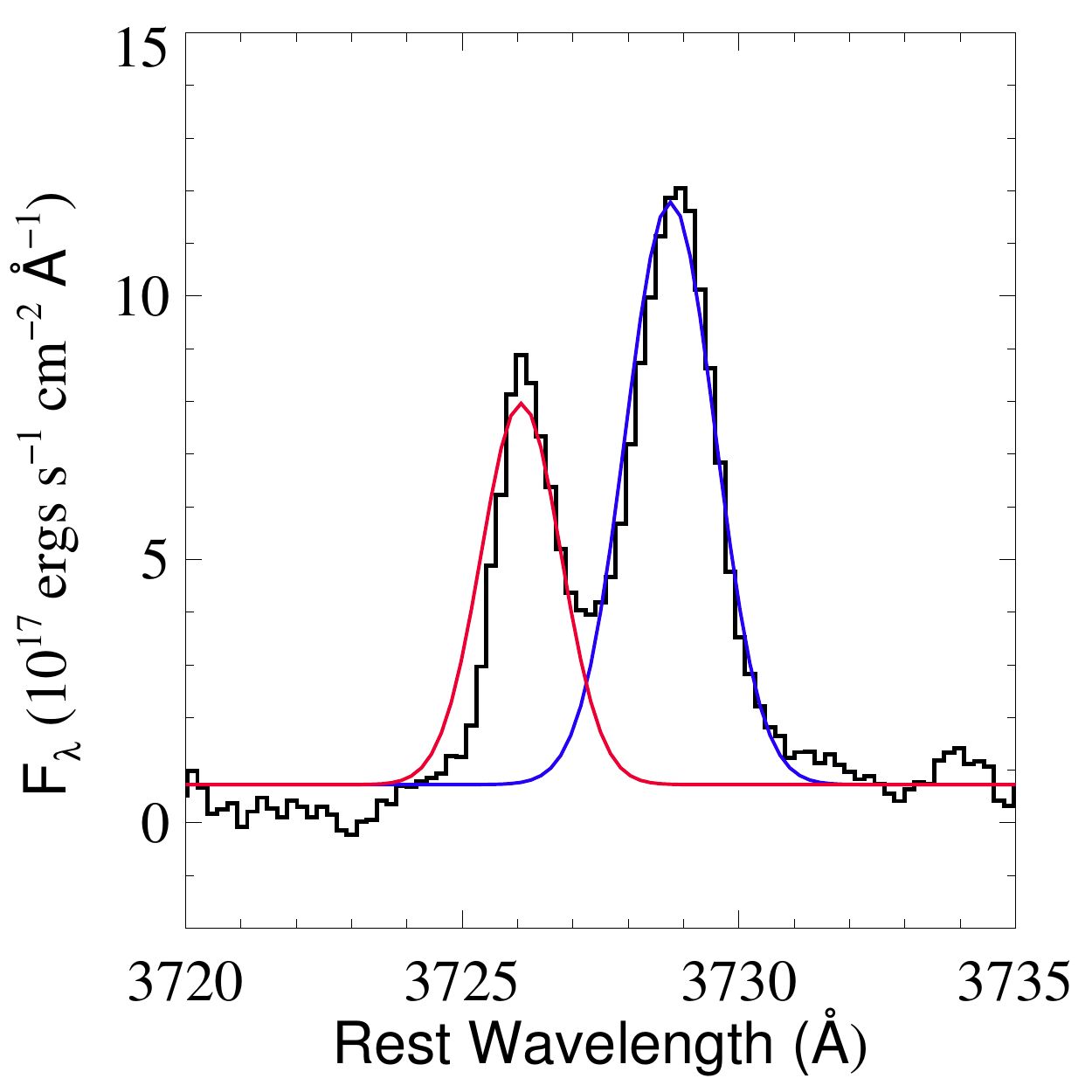}
\caption{Rest-frame resolved [OII] $\lambda\lambda$3726,3729 emission
  lines from F12$\_$ELG1, with Gaussian fits to the features
  overlaid.  The ratio of the relative strengths of the two lines
  allow the calculation of the electron density $N_e = 980 - 1280$
  cm$^{-3}$.  These data are from our Keck DEIMOS follow-up spectroscopy.
\label{oii}}
\end{figure}

The [OII]$\lambda\lambda$3726, 3729 doublet is spectrally-resolved 
in our Keck DEIMOS data, and this is
useful as these are density sensitive lines (Fig.~\ref{oii}).  The
intensity ratio is measured to be 1.44, which for a typical
temperature of T = 10,000-20,000 K equates to $N_e = 980 - 1280$
cm$^{-3}$.  This is a factor of $\sim$3 higher than the star forming
galaxies in the \citet{Kobulnicky:04} sample with 0.3$<$$z$$<$1.0, and
is consistent with ELG samples at $z$$\sim$2 \citep{Lehnert:09,
  Hainline:09}.  These, and all other fluxes in this paper are
corrected for underlying stellar absorption, extinction by dust, and
cluster magnification as described in \S4.3 and \S5.

\subsection{The SED, Underlying Stellar Absorption, and Reddening}

The hydrogen Balmer features consist of both a nebular emission and a
stellar absorption term.  The emission lines tend to weaken towards
higher level transitions owing to the sum of a rapidly decreasing
nebular emission line strength with decreasing wavelength and an
equivalent width of the stellar absorption component that at best
increases only modestly with wavelength.  To make a correction for the
underlying stellar absorption we compute an intrinsic SED model
template and then subtract it from the galaxy spectrum in a manner
similar to that of \citet{Tremonti:04}.  We note that as our object is
rare for showing H$\beta$ through H8 all in emission, the underlying
stellar absorption is not expected to be a major contaminant.

Our optical through infrared photometry is used to construct the
matching SED model.  We delens the photometry using a magnification
factor of 4.5. The observed (unlensed) photometry of F12$\_$ELG1 is as follows:
$U$=21.05$\pm$0.16, $B$=21.22$\pm$0.17, $V$=21.08$\pm$0.15,
$g^{\prime}$= 21.06$\pm$0.08, $r^{\prime}$= 20.87$\pm$0.01,
$i^{\prime}$= 20.56$\pm$0.01, $z^{\prime}$= 20.44$\pm$0.01,
$P_{3.6}$=20.11$\pm$0.07 and $P_{4.5}$= 20.46$\pm$40.08.  Our
measurements of the Balmer decrement allow a bracketed range of model
stellar ages 6$<$$t$$<$100Myr (see \S4.2).  This constraint is in
keeping with our detection of higher level Balmer emission lines, as
population synthesis models for star-forming galaxies predict for
these features an increase with evolution up to 500 Myr
\citep{GonzalezDelgado:99}.  Our model fitting approach allows for a 
range of metallicities and assumes a single starburst model with a 
range of decay rates $\tau$ and a star formation rate that depends 
exponentially on $\tau$ as follows: $SFR(t)$$\propto$$exp(t/\tau)$.
  A summary of our parameter ranges are as follows:

\begin{itemize}
\item $\tau$ = 0.1, 0.2, 0.3, 0.5, 1 and 5 Gyr 
\item $t$      = 6 - 100 Myr 
\item $Z$ = 0.2$Z_{\odot}$, 0.4$Z_{\odot}$, 1$Z_{\odot}$, 2.5$Z_{\odot}$ 
\end{itemize}

For each fixed $\tau$, 
$t$, and $Z$, two parameters are fit to the data: color excess $E(B-V)$, 
and stellar mass $M_*$.
A suite of models are constructed over the 
allowable parameter space, and each model is corrected for dust 
extinction.  We compute synthetic photometry for comparison with the 
observed photometry until the lowest value of reduced $\chi^2$ is 
obtained, as is described in detail in \citet{Frye:08}.  Our best fit model
estimate yields $E(B-V) = 0.45$ and $M_* = 2 \times 10^9 M_{\odot}$
for a young stellar age of $t = 8$ Myrs, $Z=0.4Z_{\odot}$ and $\tau =
5$ Gyr.  After redshifting and binning to the correct spectral
resolution, we measure a correction in rest-frame equivalent width of
$W = 1.5$\AA\ for the H$\beta$ line.  All Balmer emission lines
include this correction for underlying stellar absorption.  
Our
measured emission line rest equivalent widths are too low to affect the SED fits,
in contrast to the large equivalent widths found in \citet{Atek:11}.
Note although $t$
is less than the estimated galaxy crossing time by a factor of $\sim$2,
we find the BC03 models to provide an adequate fit to the data for our
purposes of determining estimates for stellar absorption.  

 The attenuation of the intrinsic light due to dust is calculated from
 a standard curve, which for a starbursting galaxy is
 given in \citet{Calzetti:00}, their Eq.~4.  Dust extinction is
 calculated using the Balmer decrement method, whereby a pair of
 emission lines with a well-defined intrinsic ratio from atomic theory
 such as Balmer lines is compared with the data.  We use 
 H$\beta$ and H$\gamma$ from our Keck dataset, and attribute
 any deviation from the intrinsic value to dust.  After correcting for
 the underlying stellar absorption, we measure $f(H\gamma)/f(H\beta) =
 0.321$.  For an intrinsic ratio of $f(H\gamma)/f(H\beta) = 0.469$
 given by \citet{Osterbrock:89}, we compute $E(B-V)_{gas}$$\approx$
 0.78.  This value is different from the color excess measured by SED
 fitting of $E(B-V)=0.45$.  Some of this discrepancy is owing to
 incomplete areal coverage of the galaxy image.  We adopt the more
 general value measured from the SED modeling for this study.

\subsection{Spectral Classification}

The spectral classification of ELGs at low redshift can be determined
from emission line diagnostic diagrams such as the classical Baldwin,
Phillips \& Terlevich (BPT) diagram \citep{Baldwin:81}. The BPT
diagram is based in part on H$\alpha$ and [N II] lines which are
redshifted out of the optical passband for $z \geqslant 0.4$.
Classification systems based on bluer emission lines are also
well developed \citep{Lamareille:04, Lamareille:09, PerezMontero:09,
  Lamareille:10,Rola:97, Marocco:11}.
\citet{Marocco:11} revise the blue emission line scheme
 of \citet{Lamareille:10}.  Their samples are derived from the Sloan
 Digital Sky Survey (SDSS) and do not include a correction
 for dust extinction.

We  classify F12$\_$ELG1 for  our ACS grism (G800L)  dataset for  which  
we have complete coverage of the emitting  line region.  We correct 
all fluxes for underlying stellar  absorption and dust extinction and  
sum up the flux over all three  ELSs.  Under the \citet{Marocco:11} 
scheme, which involves the line ratios $f$([OIII])/$f$([H$\beta$]) and
$f$([OII])/$f$([H$\beta$]), F12$\_$ELG1 is situated in the 
intermediate region that includes both star-forming galaxies (SFGs) 
and Seyfert 2 objects.  In a different classification  scheme, 
\citet{PerezMontero:09} use also the VVDS samples but instead  
consider the ratios of $f$([OII])/$f$(H$\beta$) and $f$([Ne III])/$f$(H$\beta$)  
after accounting for dust extinction.   With this diagnostic set, 
F12$\_$ELG1 is placed in a region that  includes  SFGs and  the 
uncertainty  domain between SFGs and AGNs.

The detection of additional spectral features similarly suggests that F12\_ELG1 is 
in a position intermediate between star forming and AGN source types.
We detect [CIII]/[CIV]$\lambda$4650 commonly seen in AGN and measure a flux ratio 
of [NeIII]/H$\beta$$\approx 0.6$, a value consistent with excitation by a
hard ionizing source \citep{Osterbrock:89}.   At the same time we fail to detect
[NeV]$\lambda$3426 and [HeII$\lambda$4686], two emission lines typically
associated with AGN, although the signal-to-noise is poor at the expected 
position of [NeV]$\lambda$3426 owing to its placement on top of a strong 
skyline.  Note the AGN diagnostic line [NeV]$ \lambda3346$ is blueward of our passband.  As for the
common nebular lines, the flux 
ratio $f$([OIII])/$f$(H$\beta$) is low in the center, a trend that runs contrary 
to the expected behavior of a central AGN (Fig.~\ref{figbigarc}).  Interestingly, 
the Balmer decrement is at best only marginally detected, which when used 
as an additional diagnostic places this object as an SFG for any value of [OII] 
and [NeIII] \citep{Marocco:11}.  We conclude that there is at least a strong star 
forming component to F12$\_$ELG1, and that the AGN interpretation cannot be 
ruled out.  We will use spatially-resolved spectroscopy to address the possibility 
that this object supports AGN-like activity in \S4.7.

\subsection{Gas-phase Oxygen Abundance}

The gas phase oxygen abundance is estimated from the metallicity
sensitive rest-frame optical emission lines.  The direct measurement
from the [OIII]$\lambda$4363 line typically seen in metal-poor
galaxies is not detected in any of our spectroscopy.  The indirect
measurement using emission line ratios involving H$\alpha$ is
redshifted out of our passband.  From the available emission lines we 
can compute $R_{23}$=($f$([OII]$\lambda3727$)+$f$([OIII]$\lambda\lambda$4959,5007))/$f$(H$\beta$) and 
$O_{32}$=($f$([OIII]$\lambda$4959)+$f$([OIII]$\lambda$5007))/$f$([OII]).
 $R_{23} = 4.6^{+2.3}_{-1.5}$, and $O_{32} = 0.96^{+0.39}_{-0.28}$
for the G800L dataset and after first correcting all line fluxes for underlying stellar absorption, reddening and cluster magnification.  The 
uncertainties reflect 1$\sigma$ errors in the noise and continuum placement.

The calculation of gas-phase oxygen abundances from $R_{23}$ is
complicated by its double-valued behavior, with a given value
representing either the metal-poor lower branch, the metal-rich upper
branch, or a transition at 12 + log (O/H) $\approx$ 8.4.  We measure
values for several empirical calibrations.  Our value for $R_{23}$
yields an abundance of 12 + log (O/H) = 8.1 (lower branch) and 8.9
(upper branch) \citet{Kobulnicky:04}.  For \citet{Tremonti:04} as
adapted from the method of \citet{Charlot:01} we obtain 12 + log (O/H)
= 8.8 (upper branch).  For \citet{Zaritsky:94} we measure 12 + log
(O/H) = 8.8 (upper branch).  Finally, we compute 12 + log (O/H) = 7.8
(lower branch) and 8.7 (upper branch) for \citet{McGaugh:91} as given
by \citet{Kobulnicky:99}.  F12$\_$ELG1 is consistent with being on either
branch.

Of the available emission features, the secondary indicator [NeIII]
can break the metallicity degeneracy \citep{Nagao:06}. We compute
$f$([NeIII])/$f$([OII]) = $0.05^{+0.01}_{-0.07}$, which has a best-fit
polynomial correspondence value of 12 + log(O/H) = $8.8 \pm 0.2$.  
Our value is close to the solar value of 12 + log (O/H) = 8.66
\citep{AllendePrieto:02, Asplund:04}.  It also overlaps with the range of
values obtained from current metallicity history studies at $z$=1-2 of 12 + log(O/H) = 8.3 - 9.0
\citep{Lamareille:06b, Liu:08, Hainline:09}.  In turn, all values are higher
than those of direct measurements from strongly-lensed high-z galaxies 
\citep{Yuan:09, Rigby:11}.

\subsection{Star Formation Rate}
We estimate the star formation rate in two ways:  (1) by extrapolating
the intrinsic H$\beta$ line flux into an estimate of the H$\alpha$ flux, and
(2) from the intrinsic (corrected for reddening and lensing magnification)
[OII] line flux.  Although our observations do
not cover the most reliable tracer of star formation, H$\alpha$, one
can infer the H$\alpha$ line flux from the $E(B-V)$ value and the
intrinsic flux ratio between H$\beta$ and H$\alpha$
\citep{Osterbrock:89}.  We estimate the intrinsic fluxes for H$\alpha$
to be: $f$(H$\alpha)$=1.4$^{+0.53}_{-0.61}\times10^{-15}$ ergs
s$^{-1}$ cm$^{-2}$ and $f$(H$\alpha)$=2.1$^{+0.38}_{-0.33} \times
10^{-16}$ ergs s$^{-1}$ cm$^{-2}$ for the Keck and HST G800L
observations, respectively.  The SFR can be measured using the
relation starting from \citet{Kennicutt:98}: $SFR (M_{\odot}/yr)$=$7.9
\times 10^{-42} f(H\alpha)*4\pi D_L^2$ ergs s$^{-1}$.
  For the Keck observations we measure SFR
=31$^{+11}_{-13}$ M$_{\odot}$ yr$^{-1}$.  For the HST G800L observations there is
insufficient spectral resolution to detect H$\beta$ in every ELS,
making our value of SFR $\apg$ 6.5 M$_{\odot}$ yr$^{-1}$ a lower
limit.  We will adopt the value for the SFR from the Keck 
observations for this study.

 Calibration of the star formation rate using the intrinsic [OII]
 $\lambda$3727 line is more challenging, as it is more sensitive to
 reddening, metallicity and ionization parameter.  \citet{Kewley:04}
 present an algorithm that takes these dependencies into account,
 especially for high gas-phase oxygen abundances which apply to our
 case (log (O/H) + 12 $>$ 8.4). Their Equation 14 for SFR is based on
 ionization parameter and gas-phase oxygen abundance.  Using this
 equation yields SFR$_{[OII]} \approx 3$ M$_{\odot}$ yr$^{-1}$ for the
 G800L grism dataset, a value that is small compared to estimates
 based on H$\beta$ flux.  We will adopt the value for the SFR from the 
 H$\beta$ flux for this study as its relation to H$\alpha$ is better understood, 
 and plan to measure SFR directly by H$\alpha$ in grism observations with 
 VLT SINFONI in an upcoming paper.  Finally, we compute sSFR$\approx20$ 
 Gyr$^{-1}$.

\begin{figure}[h]
\includegraphics[viewport = 0 0 200 460,scale=0.65]{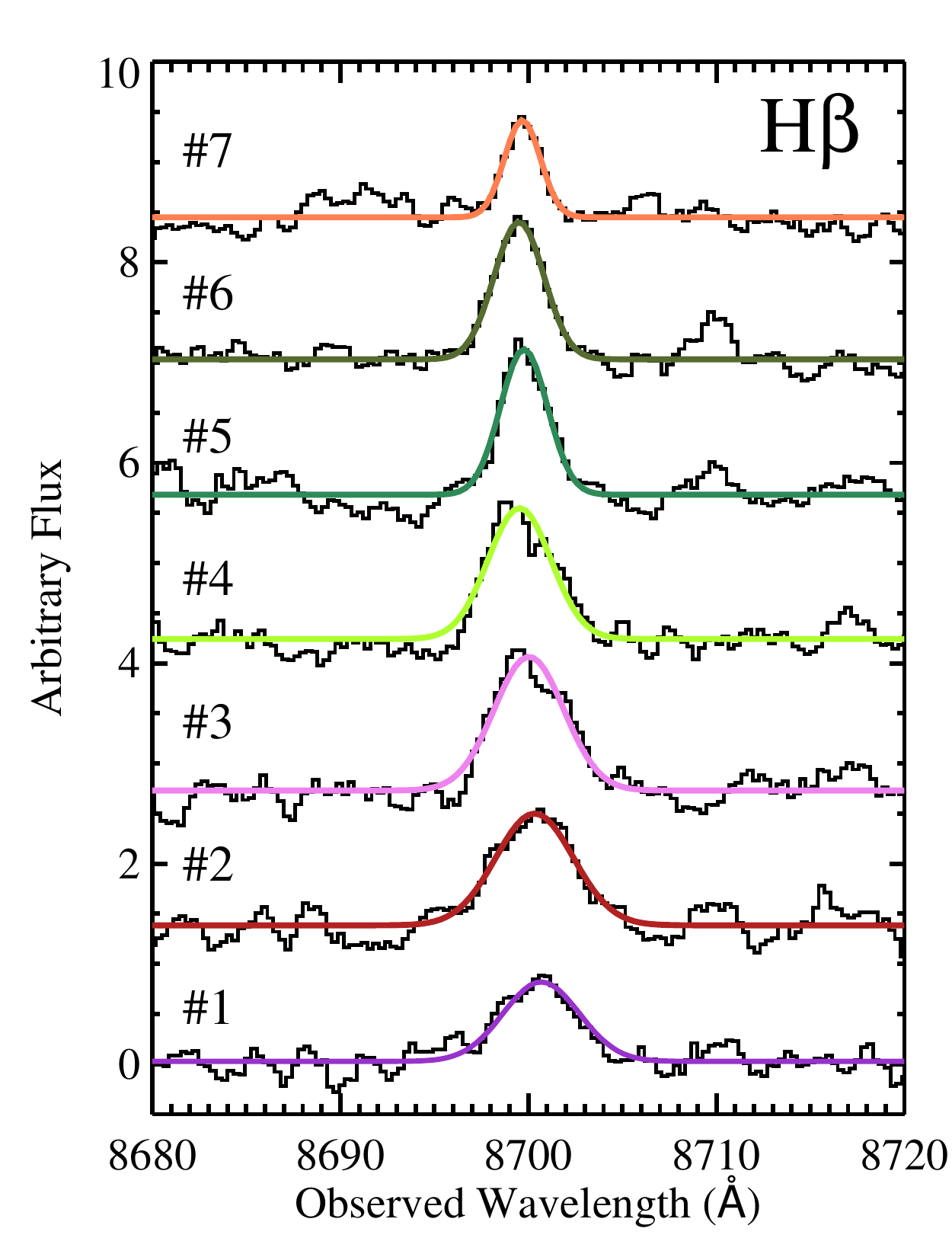}
\caption{Spectra centered on the H$\beta$ emission line are plotted for each 
of the seven slices in our spatially-resolved Keck spectrum of F12$\_$ELG1 
at $z=0.7895$.   The colors are matched to the colors in Fig.~9.
The line fits are used to determine the central wavelengths 
and FWHM's.  We find a small Doppler shift in the line centroid across the bins. 
\label{Balmer}}
\end{figure}

\begin{figure}[h]
\includegraphics[viewport = 10 50 200 500, scale=0.45]{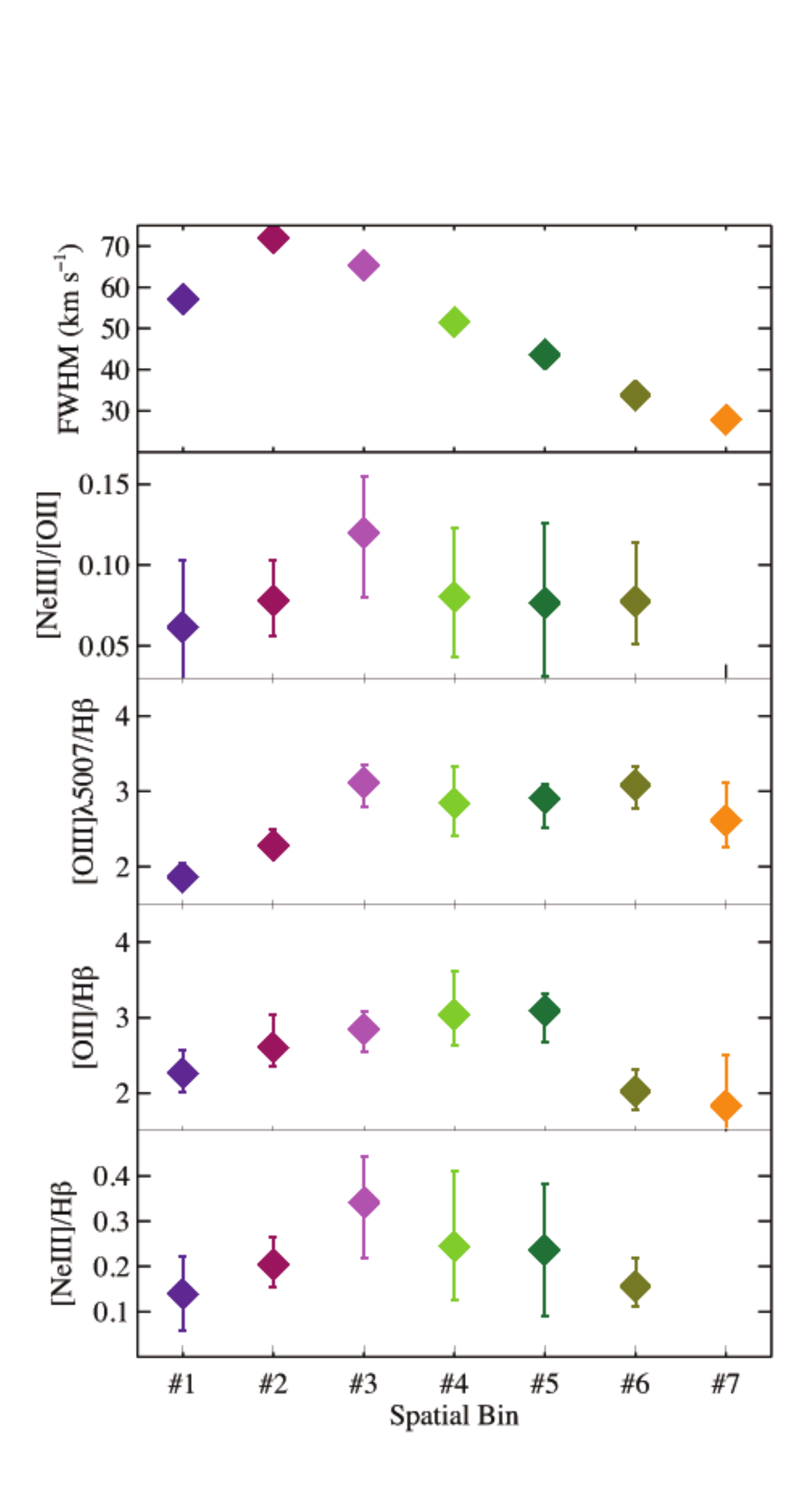}
\includegraphics[viewport =   75 -400 700 85,scale=0.95]{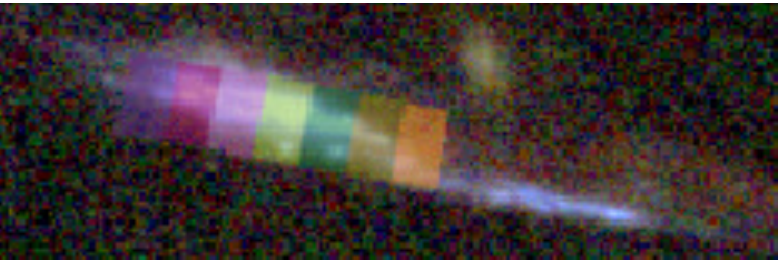}
\caption{$gri$ image of F12$\_$ELG1 at $z=0.7895$, with trends in the FWHM
of a strong emission line and
metallicity dependent line ratios shown along the substructure of this giant arc. 
Each spatial bin covers 
0.6$^{\prime \prime}$ or $\sim$1 kpc (intrinsic).  The FWHM peaks at Bin 2, 
which is taken to be the dynamical center.  The metal 
fraction peaks in Bin 3, which is offset from the dynamical center.  The metal
line ratios in Bin 3 admit the possibility of an AGN.  This asymmetrical AGN-like
behavior may be a result of shocks in the direction of the long extended tail.
 \label{trends} }
\end{figure}

\subsection{Spatial Trends in F12$\_$ELG1}

The image of this giant arc is noticeably bisected into a broad, line-emitting
clump  and a long extended tail.  Line emission is detected only in the clump.  
We have spatially-resolved
spectroscopy with sufficient angular resolution to subdivide the line
emitting clump into seven contiguous bins of 0.6$^{\prime \prime}$
each, or equivalently 1kpc each after correcting for a magnification
factor of 4.5.  Each bin is assigned a different color.
The bins have the following assignments: Bin
1 (purple), Bin 2 (rose), Bin 3 (lavender), Bin 4 (light green), Bin 5
(dark green), Bin 6 (olive), and Bin 7 (orange).  We extract the seven
spectra and compute line fluxes, line centroids and velocity widths
for common emission lines.  A sample stack plot centered on the
position of H$\beta$ in particular is shown in Figure ~\ref{Balmer}, and an 
image of F12$\_$ELG1 with the seven bins overlaid appears in the top panel
of Figre~\ref{trends}.

The velocity dispersion $\sigma$ is derived from the FWHM of the H$\beta$
line and based on the assumption that the line width is set by the spread of velocities
of the line emitting gas about the local central mass.  The
instrumental profile is subtracted off of the FWHM in quadrature, and
then $\sigma$ is computed as follows: FWHM/2.355*c/$\lambda_{obs}$.
There should be a peak at the galaxy's center of mass 
if the object is supported by a central mass pulling on the stars.
For F12$\_$ELG1 the peak is at Bin 2, making this location the likely
location for the galaxy nucleus (Fig.~\ref{trends}).
The uncertainties in $\sigma$ come from continuum fitting and are less than
the size of the plotting symbols.
 %
 It is interesting that the H$\beta$ line centroid {\it blueshifts} relative 
 to the mean value along the galaxy, from $\Delta v = +35 \pm
5$ km s$^{-1}$ for bin 1 to $\Delta v =-13 \pm 5$ km s$^{-1}$ in bin
7.  These results are similar to what is also seen across the lensed
galaxy Lens22.3 at $z=1.7$ \citep{Yuan:09}.  From the velocity
dispersion one can compute a rough value for the dynamical mass of
$\sim2 \times 10^{9}$M$_{\odot}$.  This value is consistent with our
estimated stellar galaxy mass from SED-fitting.  We measure $M_B=-22.2$,
and in turn obtain an estimate of the mass to light ratio of M/L
$\approx$1.  
Our low measured M/L and stellar mass is consistent also with our
stellar age by SED fitting of $t=8$Myr.  This is similar to 
the work of
\citet{vanderWel:05} who find a low  M/L and low 
stellar mass at $z$$\sim$1 to be correlated with a young stellar age.  

The trends of various metallicity-dependent line ratios across
the galaxy are also shown in Figure ~\ref{trends}.  We retain the same bin
coloring scheme and apply the same modest correction for stellar
absorption and extinction to each bin as computed in \S4.3.  While
the peak in the FWHM is at Bin 2, 
there is a temptation to find a peak over the metal line ratios at Bin 3, and
a drop off to the left-hand-side (Bins 1 and 2).  We derive below a
number of interesting results by a comparison of values in Bin 3 with
the mean of Bins 1 and 2 (Bins 1-2).  By \citet{Marocco:11} Bin 3
encompasses the composite star forming galaxy (SFG)/Seyfert 2 and
Seyfert 2-only regions of their $f$([OIII]/H$\beta$)
vs. $f$([OII]/$H\beta$) H II excitation diagram.  Meanwhile Bins 1-2
yield a different result, and reside entirely in the SFG/Seyfert 2
region.  By the approach of \citet{PerezMontero:09}, which includes
[NeIII], Bin 3 plus uncertainty region lies in the transition region
between SFGs and AGNs while Bins 1-2 plus uncertainty region lies
entirely {\it inside} the SFG region.  We conclude that the disky Bins
1-2 are consistent with an SFG while Bin 3 exhibits AGN-like behavior.  
These strong metal line ratios that are driven by a hard ionizing source
and are offset by $\sim$1kpc from the peak of the velocity dispersion appear
to be best explained by shocks in the direction of the galaxy tail (Bins 4-7).  

The values for [OIII]/H$\beta$ are large all across the galaxy, and
are also higher in Bin 3 compared to Bins 1-2.  In Bin 3 we measure
log($f$([OIII])/$f$(H$\beta$)) = $0.49 \pm 0.04$.  There is only a
hint of a trend in the behavior of [NeIII]/H$\beta$, but it is worth
noting that [NeIII]/H$\beta$ is high in the central three bins, and
for Bin 3 in particular the value is consistent with 0.4, a value
commonly associated with AGN \citep{Osterbrock:89}.  It is tempting to
say the disky Bins 4-7 also show a decline from Bin 3, and if so then
the various line ratios are not falling off as rapidly as in Bins 1-2.
We speculate that there is a moderately elevated metal fraction in
Bins 4-7 that may be a result of increased star formation activity in
the direction of the galaxy tail, indicating possible harassment in the galaxy's
star formation history.

If Bins 4-7 indeed lead towards a galaxy tail, then one can ask the question
of which are the likely culprits for past interactions?  
F12$\_$ELG1 has neighbors roughly 
centered in redshift at the systemic redshift of $z$=0.7895.  There are 
seven galaxies with spectroscopic redshifts of 0.7625$<$$z$$<$0.8175 
distributed over a field in the image plane of $\approx 3.5$ arcmin on a side, 
which for a mean magnification of $\sim4\times$ yields a physical size in the 
source plane of 0.8 Mpc.  This structure or filament may contribute to the 
compound lensing effect at the $1\arcsec$ level that is important for doing 
precision cosmology with clusters \citep{Jullo:10}.

\subsection{Evolution of F12$\_$ELG1}

In sum, with its high sSFR corresponding roughly to a cold gas
fraction of $\sim$0.7 \citep{Reddy:05}, its low mass, and presence of
many Balmer series emission lines, this strongly-lensed but otherwise
ordinary galaxy would appear to be caught at the beginning of a major
burst of star formation.  This is consistent with the picture of
\citet{vanderWel:11} in which $\sim$1/2 the stars in a typical field
galaxy are formed in only $\sim$2-3 bursts that produce the stellar
population in a $M_*$=10$^9$M$_{\odot}$ by $z$$\sim$1.  F12$\_$ELG1 also
looks similar to the objects in \citet{Kriek:09} that are AGN hosts of
size $\sim$1 kpc.

F12$\_$ELG1 has similar ionization properties compared to other 
$z$=1-2 star-forming galaxies, and as a group such objects have 
elevated ionizations compared to the local galaxy population.  It is 
thought that galaxy feedback must play a role in understanding 
these differences.  One possible explanation is that many intermediate
redshift galaxies may harbor weak AGNs \citep{Wright:10, Trump:11}.
F12$\_$ELG1 has elevated metal-line ratios consistent with an AGN 
that are interestingly offset from the dynamical center of the galaxy.  
Given this extended tail in 
the same direction, we interpret the asymmetrically-situated AGN-like 
region of this galaxy as shock excitation possibly as a result of a past 
galaxy interaction.  The line ratios decrease towards the outer disk on 
both sides of the peak in this one galaxy, similar to the spatially-resolved
spectroscopy of another giant arc at intermediate redshift, the `Clone' 
\citep{Jones:10b}.  By contrast, in a sample of 50 intermediate redshift 
ELGs with integral field unit spectroscopy a significant fraction showed 
the opposite trend \citep{Queyrel:11}.

\section{Cluster Lensing}

Abell 1689 is one of the best studied clusters in the literature, with
several published lens models \citep{Broadhurst:05, Limousin:07,
  Halkola:06, Leonard:07, Zekser:06, Coe:10}.  In this work we adopt
the model presented in \citet{Broadhurst:05}, which is constructed
from 30 multiply-imaged galaxies with supporting multiband photometry
of the field and spectroscopy of a representative sample of the arcs.
The \citet{Broadhurst:05} model compares favorably with measurements
using an independent model from the literature \citep{Limousin:07}.

\begin{figure}[h]
\includegraphics[viewport=0 0 250 330,scale=0.53]{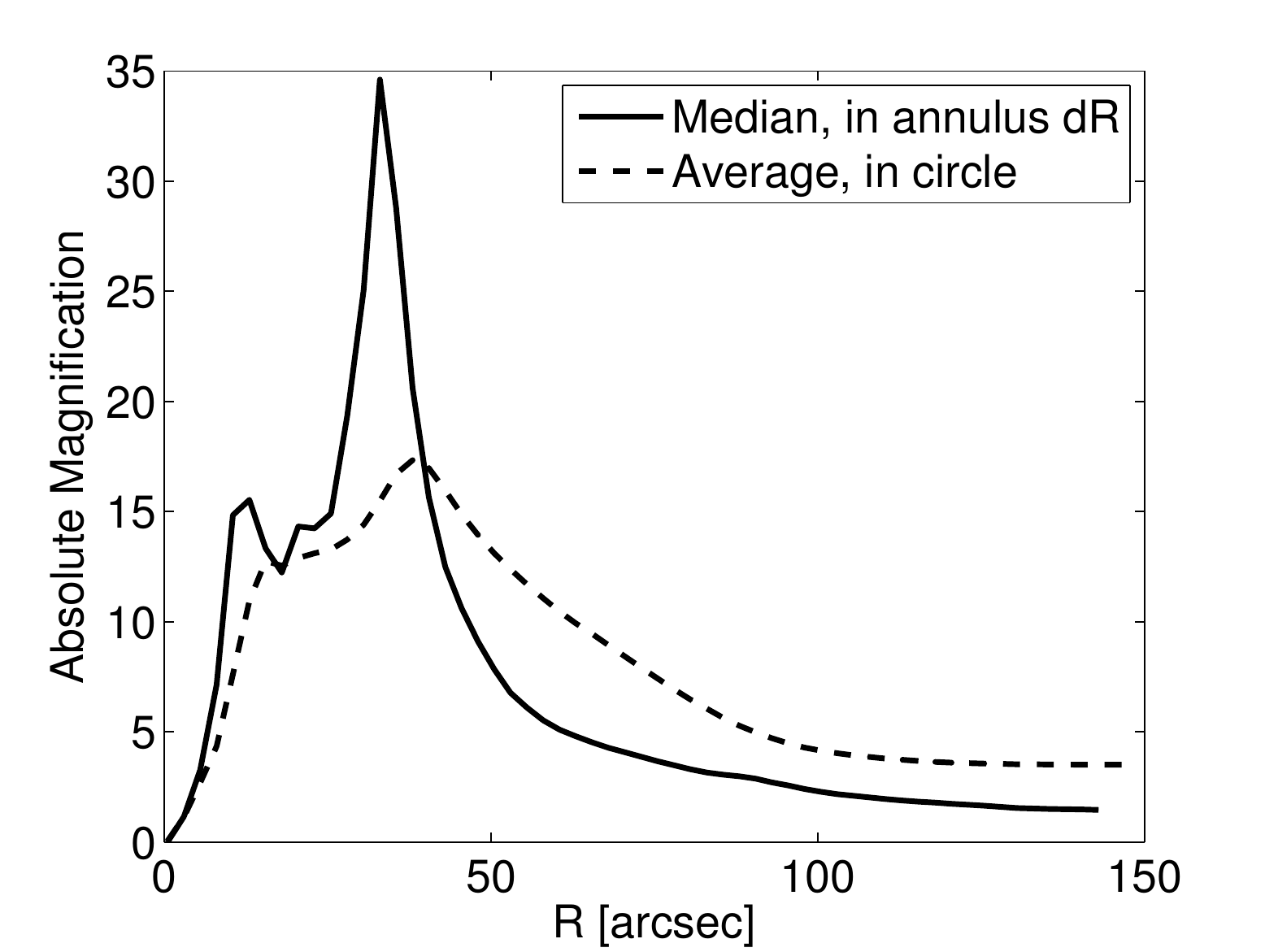}
\caption{Radial magnification profiles for the average magnification inside
a circle of radius $R$ (dashed line), and the median magnification in an annulus
of 2.5$^{\prime\prime}$ about $R$ (continuous line) assuming in each case a 
mean source redshift of $z=0.8$ as drawn from the literature \citep{Frye:07}.  
The primary and secondary peaks appear at the locations of the tangential 
and radial critical curves, respectively.  The net advantage of this lensed field is
a factor of $\sim$4 integrated over a large $R=150$$\arcsec$ (0.5 Mpc) radius.
\label{figmagnif}}
\end{figure}

We use our lens model for A1689 to construct an azimuthally-averaged 
magnification profile to predict source magnifications ($\mu$) owing to strong 
lensing.  We measure the magnification straightforwardly by tabulating a grid 
from the best-fitting magnification map 
that spans roughly the ACS field of view (190$\arcsec$ on a side).  
Our resulting magnification profile is shown in Figure~\ref{figmagnif}, 
corresponding to the measured mean redshift of background objects in the 
field drawn from the literature \citep{Frye:07} and references therein. 

The median magnification ($\mu$) peaks at $\mu$=35 and has a long tail 
extending towards large radial distances. We report the median magnifications 
rather than the mean magnifications so as not to include the high model 
magnifications exceeding $\mu$=100.  The magnification of both curves is 
high outside the tangential critical curve of 50$\arcsec$, even out to large radii of
$R$=150$^{\prime\prime}$ ($\approx$ 0.5Mpc), where the median absolute
magnification is $\mu$$\approx$4.  Given this magnification profile, all the 
sources in our catalog and behind the cluster are likely magnified.  The 
magnification is computed in detail for two giant arcs of interest below.

\begin{figure}
\includegraphics[viewport = -405 343 250 345,scale=0.33]{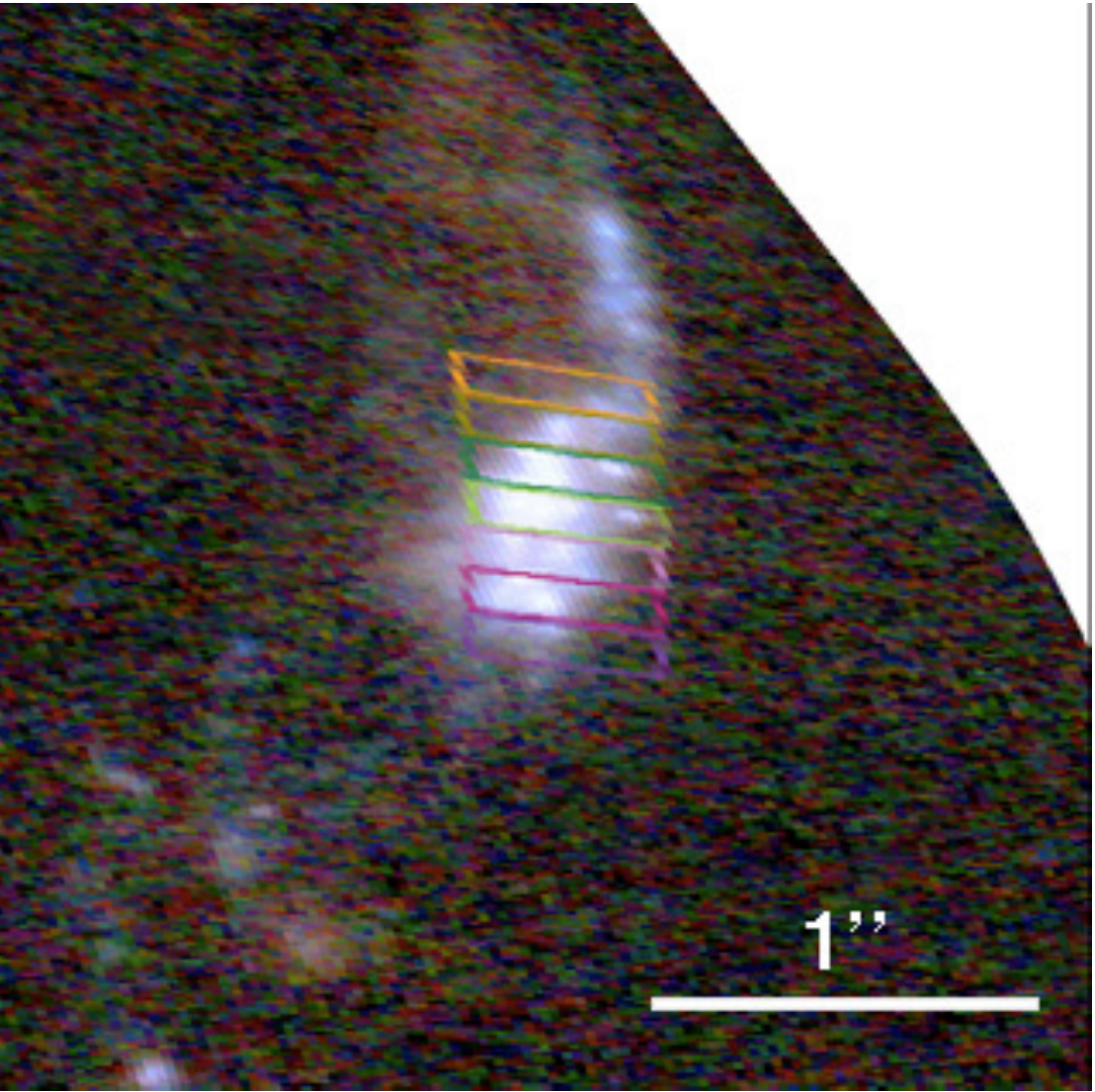}
\includegraphics[viewport = 65 25 250 400,scale=0.325]{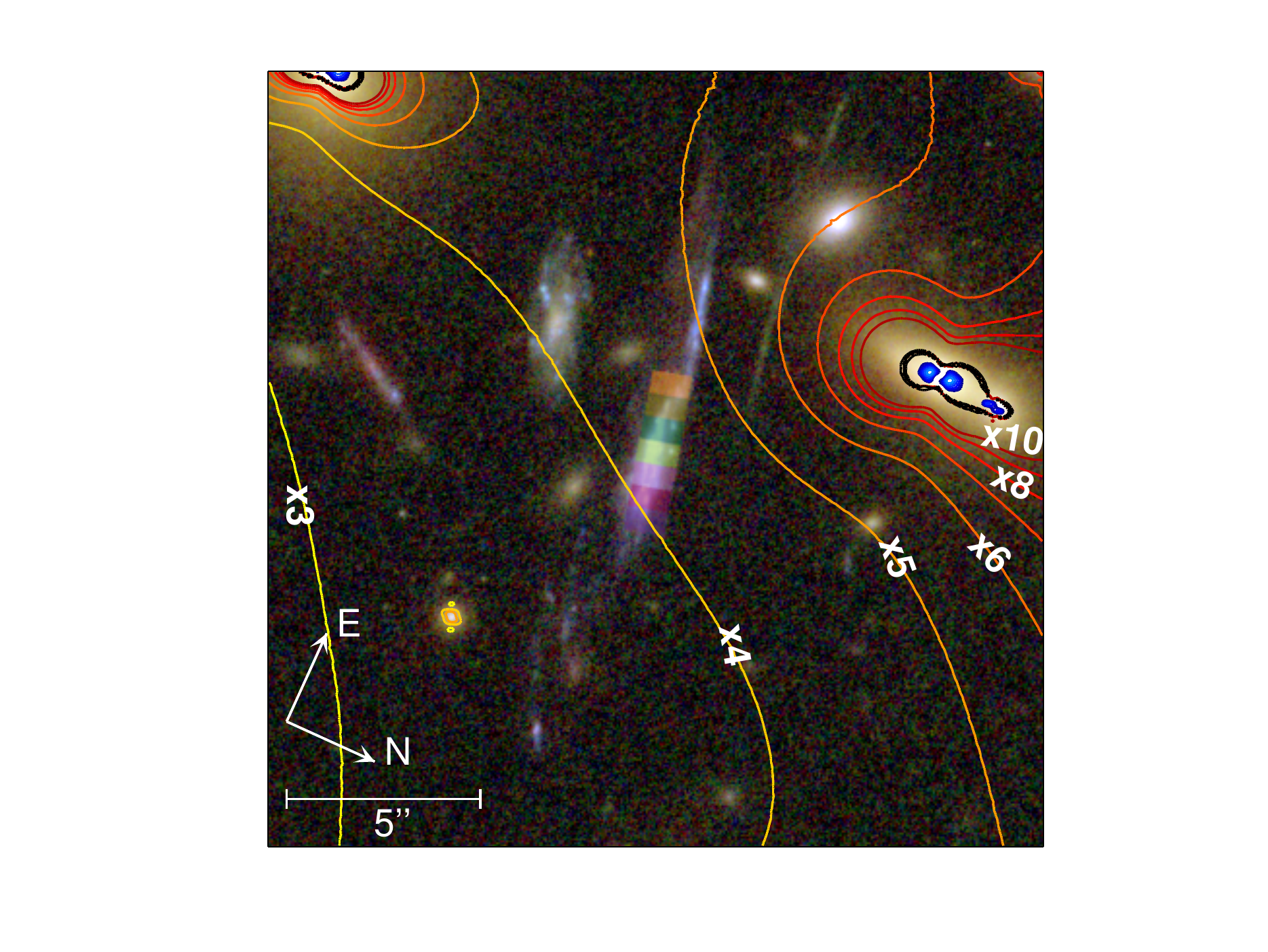}
\caption{Left-hand side:  two-dimensional magnification map of the lens 
A1689 (contours) overlaid on the $gri$ image of the blue giant arc at $z=0.7895$.  
The linear magnification varies differentially from $\mu = 4-5$ along the
arc, enhanced locally by the nearby cluster elliptical at right.  Seven 
non-overlapping bins of substructure are overlaid as colored rectangles 
in both the image and source planes.  The bins are oriented and have a 
size corresponding to the position angle and slit-width of the observations.
Right-hand side:  the reconstructed source plane image yields a single nucleus with
an extension that may be a tidal tail.  \label{rainbow}}
\end{figure}

The magnification contours for F12$\_$ELG1 are laid down onto the HST
$gri$ image in Figure~\ref{rainbow}.  The contours increase in the
direction of the two massive cluster ellipticals appearing just above
the ``$\times$10" magnification label and in the upper left-hand
corner of the image.  We compute a magnification that increases from a
factor of four to five along the long axis of the arc, with a mean
value that we use for this paper of $\mu = $4.5.  This value
supercedes a previous measurement \citep{Frye:02}, as this one is
derived explicitly from the lens model.  No counterarcs are predicted
for this arc which is not situated close the expected positions of the
critical curves.  The reconstructed source plane image 
shows one bright compact region of size $\sim1\farcc$ and an
extended tail.

\section{Other Field Galaxies}

\subsection{Double-peaked H$\alpha$ in F12$\_$ELG2}

\begin{figure}
\includegraphics[viewport= 0 0 200 360,scale=0.65]{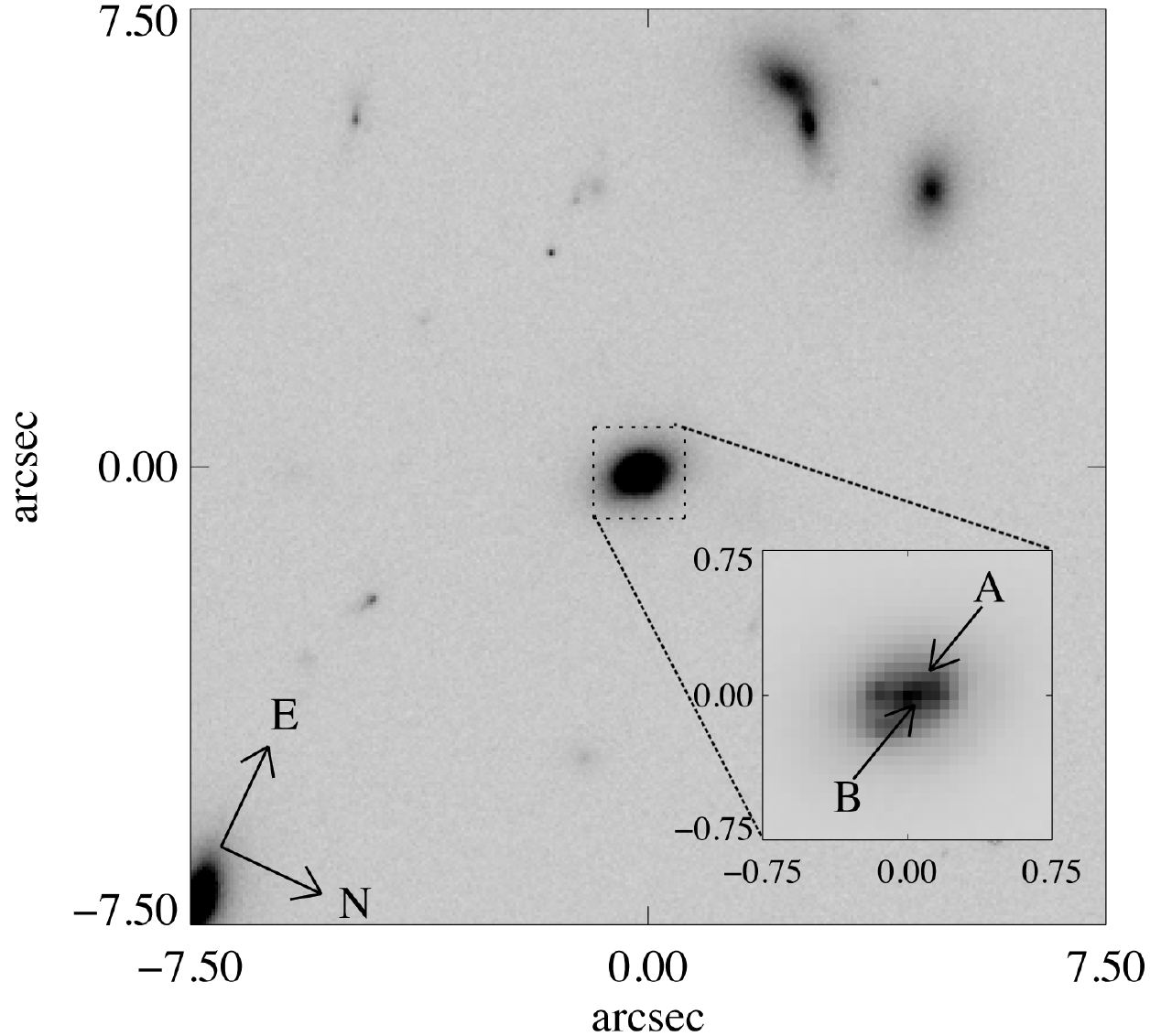}
\caption{Image in $i_{775}$ of F12$\_$ELG2 (center).  
The inset image is a close-up of the region indicated
by the dotted lines that contains two emission line sources (ELS 20004 `A' 
and ELS 11085 `B').  ELSs A and B have a close angular separation of 
$\theta$$\approx$$0.2\arcsec$,  and other knots are also detected in the vicinity
of this elliptical galaxy.  We show the grism spectroscopy in Figure~\ref{2humps}.
\label{2humpim}}
\end{figure}

\begin{figure}
\includegraphics[viewport=  -30 -35 520 750,angle=90,scale=0.35]{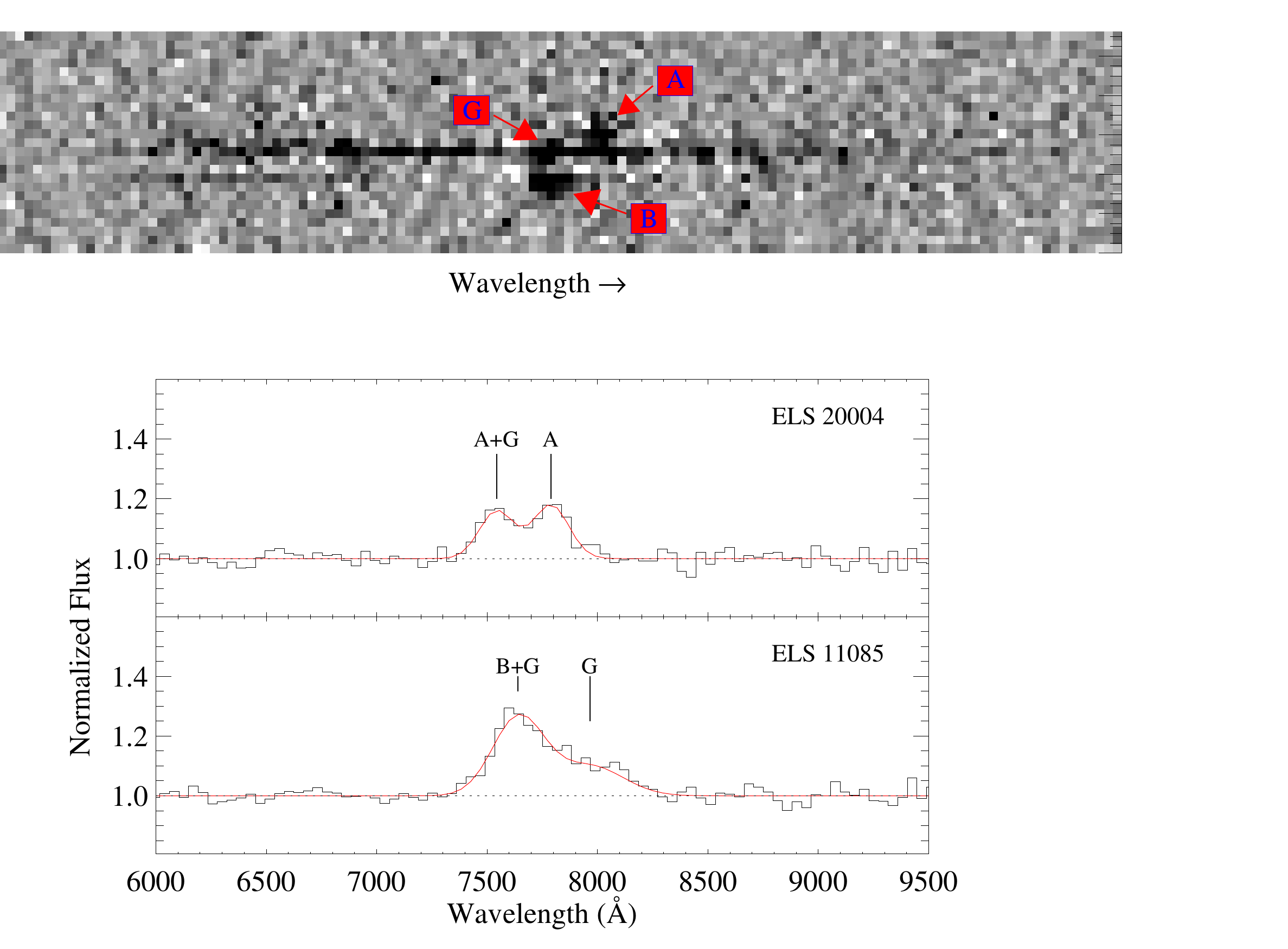}
\caption{Spectra of F12$\_$ELG2, which contain new ELSs at close angular
  separations: ELS 20004 `A,' ELS 11085 `B,' and 'G' (see
  Fig.~\ref{2humpim} for the image).  From the line placements and
  separations we infer there to be one emission line for each of the
  three components.  The spectrum in the top panel is remarkable for
  showing a double-peaked emission line profile with a large
  peak-to-peak velocity difference of $\sim$9600 km s$^{-1}$, and line
  strengths of roughly equal amplitude.  Component G shows underlying
  stellar continuum.  Most likely A and G are at $z$=0.532 while A is at
  $z$=0.560, although additional data are required to confirm these 
  line identifications for this complex system of ELSs.
\label{2humps}}
\end{figure}

We report new ELSs detected with the ACS G800L grism
which have close angular separations.  
They are: ELS 20004 `A,' ELS 11085 `B' (Fig.~\ref{2humpim}).  We
designate this group of sources as `F12$\_$ELG2.'  Sources A and B have an
angular separation of $\approx$0.21$\arcsec$, and are resolved despite
the angular separation being slightly smaller than our predicted
spatial resolution of $R_0 = 0.25\farcc$ set by our algorithm that
extracts a minimum of five spatial pixels centered on an emission
line.  The spectrum for component A shows two emission line peaks with
similar flux amplitudes.  The spectrum for component B shows a single
emission line with an extended red tail that appears to be associated
with at least some of the same emission sources as in component A
(Fig.~\ref{2humps}).

Emission peaks B and G both have line emission at the same wavelength
(see 2d spectrum in Figure~\ref{2humps}).  In addition component G
shows extended emission, making it the source likely to be most
sensitive to the photometric redshift estimate.  The photometric
redshift for G is $z_{BPZ} = 0.480 \pm 0.145$, from which we consider the
peak `B+G' as [OIII] at $z$=0.532.  As it is unlikely for components B
and G to have the same wavelength peak but be unrelated, we take this
as the redshift for both components.  The morphology of the galaxy
nearest to the sources is an elliptical which is likely component G (see Fig.~\ref{2humpim}).
From profile fitting we do not find associated H$\beta$.
Our double-peaked spectrum (A+G) has a peak-to-peak velocity
separation of $\sim$9600 km s$^{-1}$.  From this remarkably-high value
we can rule out the interpretation of this object as two separate ELSs
close in space.  Most likely, G and B are both at $z$=0.532, while A is an 
unrelated source at $z$=0.560.  It is also possible that source A and/or B are
situated behind the elliptical galaxy, which would alter the above redshift identifications.  
Additional spectroscopy is required at a competitive spatial resolution to identify 
these emission features and objects.

\subsection{High-$z$ Population}

\begin{figure}
\includegraphics[viewport=-110 -760 250 230,scale=0.25]{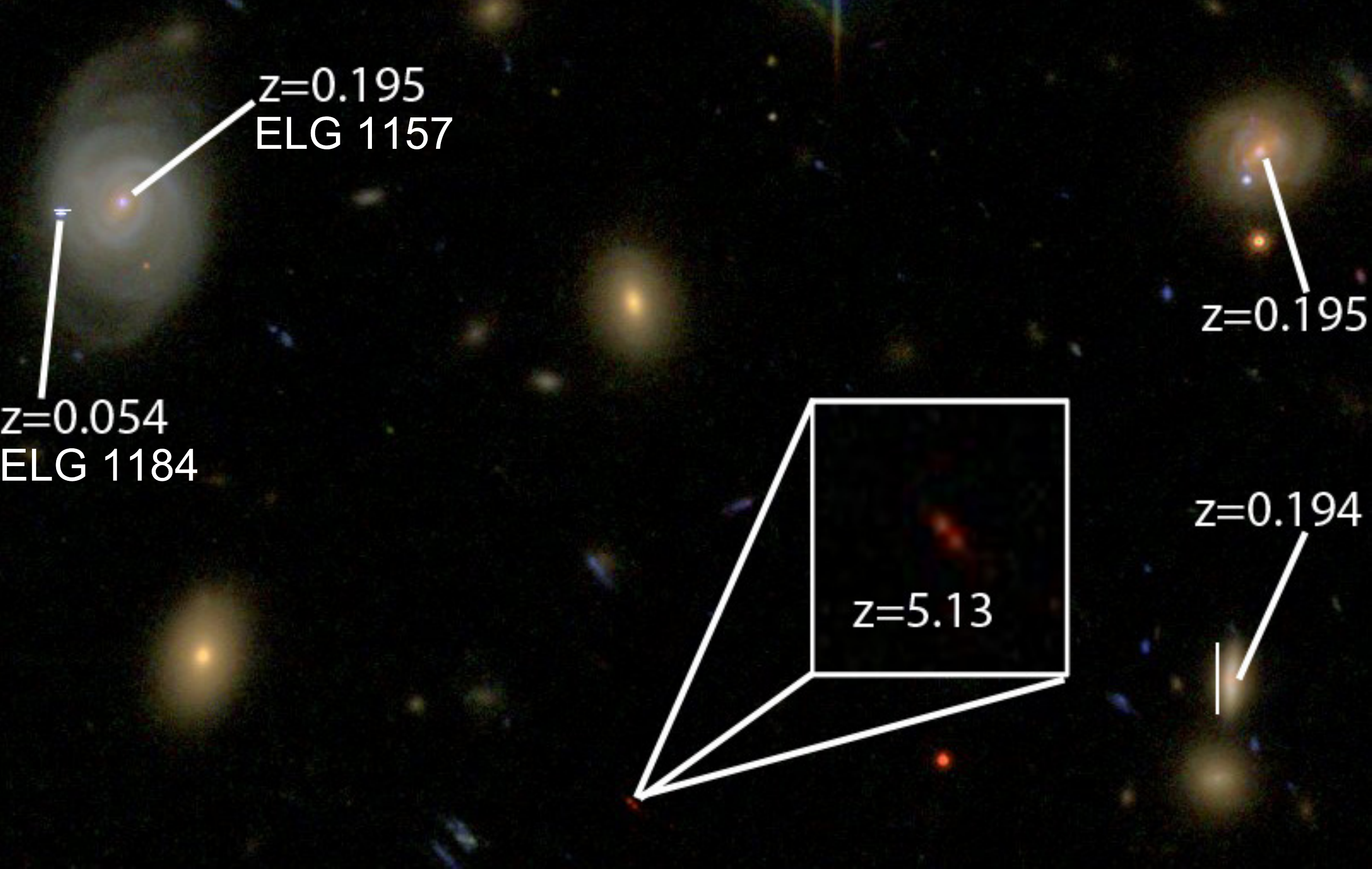}
\includegraphics[viewport=   162 -10 150 500,scale=0.64]{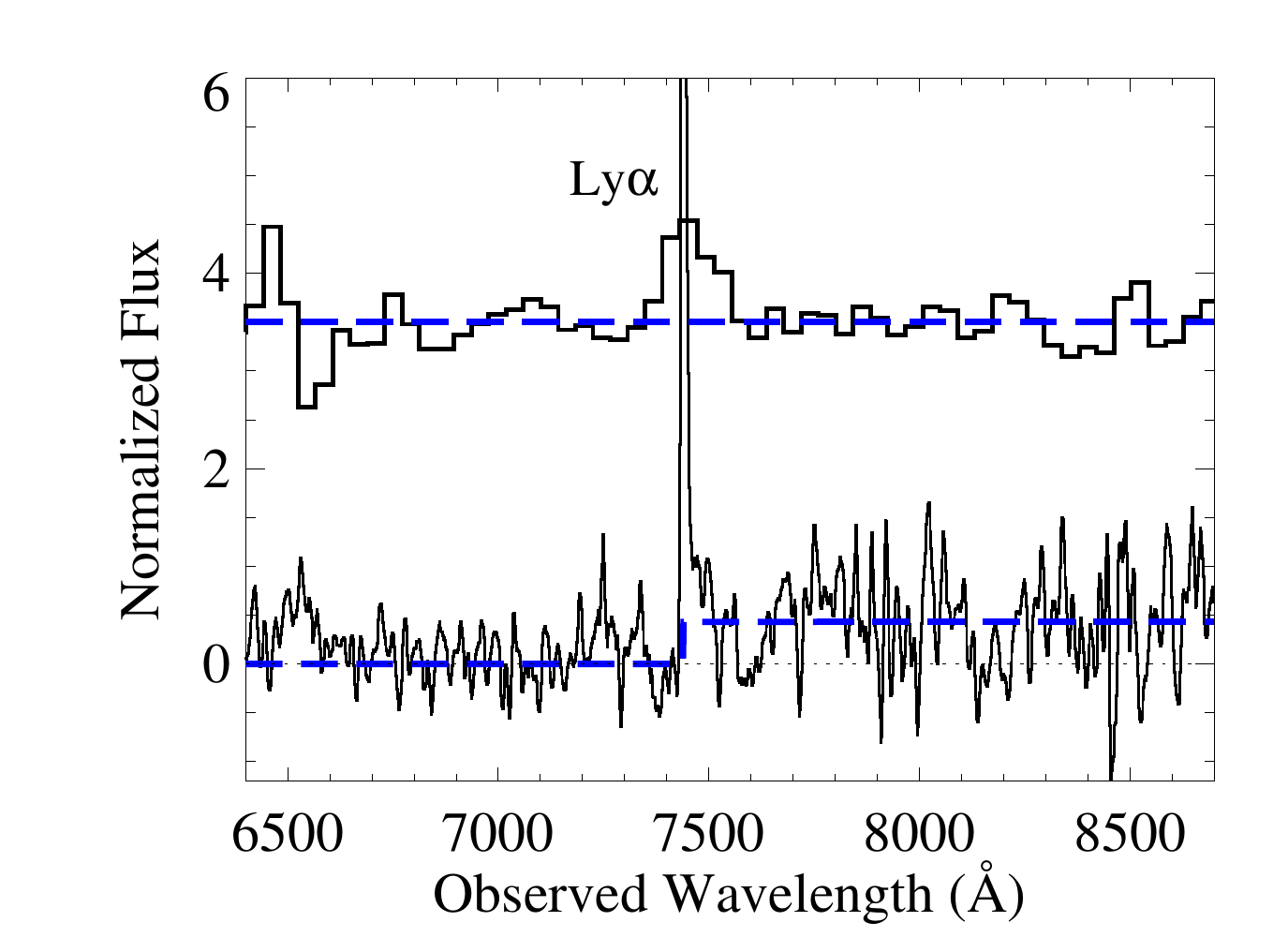}
\caption{$gri$ image and spectra of ELG 10399 at $z=5.13$
($i_{775} = 25.58 \pm 0.18$).  A Lyman-series break is clearly detected together 
with a prominent asymmetric emission line which we take to be Ly$\alpha$
from our Keck LRIS observations (lower spectrum).  Our grism results are 
also shown (upper spectrum).  Two other ELSs from 
our sample appear in the field.  
 One is  ELS 1184 at $z=0.054$ and the other is ELS 1157 at $z=0.195$,
as marked.  Spectroscopic redshifts of other objects in the 
field known from the literature are marked. \label{fighiz}}
\end{figure}

We detect the highest redshift object with a published spectrum in 
the A1689 field, ELG 10399 at $z$=5.13 \citep{Frye:02}.    Our grism spectrum is shown in the upper panel.
This arclet is faint ($i_{AB}= 25.85 \pm 0.18$) and
extremely red ($g-i$$>$4), yet is detected in
our rather shallow grism survey owing to its large rest-frame equivalent width
($W_r$=29.7$^{+13.29}_{-4.6}$), its high magnification, and its small size of
$1^{\prime\prime}$ (Fig.~\ref{fighiz}).  We compute a magnification of $\mu
\approx 4.5$, and estimate an intrinsic size for its redshift of
$\approx1$kpc.  There are no counterarcs predicted for this arclet.
We identify the lone emission line to be Ly$\alpha$ based on our
higher resolution companion spectrum taken with Keck LRIS (blue dashed
line) which shows also a Lyman-series break.  With its high rest
equivalent width of $W_r > 20$, this object would appear from the
grism spectrum alone to be a Ly$\alpha$ emitter, but our Keck spectrum
shows it to have a Lyman-break seen against stellar continuum.

Of the six galaxies in the field of A1689 with $z$$>$2.5
(Table~\ref{tablehiz}), interestingly only this one galaxy at $z$=5.13
enters into our sample.  The Sextet arcs at $z$=3.038 are not
detected, which is not surprising given that the arclet in this
sextuply-imaged system with the strongest line emission has a
Ly$\alpha$ total of $W_r$=4$^{+1.5}_{-5.0}$\AA, a value in the lowest
single percentile of our sample.  It is less well understood why the
bright arclet at $z$=4.868 is not detected.  This arclet image has
$i_{775}$=23.48$\pm$0.03, an angular size of $\sim$1.3$\farcc$ a
Ly$\alpha$ emission line with $W_r$=12.4$^{+8.83}_{-3.84}$\AA \ (the
lowest 10\% of our sample), and is situated in a relatively uncrowded
location relative to the cluster members.  Note we do not see any lone
unassociated emission lines of other faint, potentially high-redshift
objects near to the locations of the tangential critical curve.

\section{Summary and Future Work}

We have undertaken a census of emission line galaxies in the central
portion of A1689 comprising three orbits and a single pointing with
the HST ACS G800L grism.  This is the first grism survey with HST in
the field of a massive lensing cluster.  We summarize the main results
below.

\begin{itemize}

\item{We present a spectroscopic catalog (Table~\ref{ID}) which 
contains 66 emission lines in 52 emission line sources in 43 galaxies in 
this flux-limited sample with $i_{775}{^{<}_{\sim}}27.3$.  Three-fourths of 
the galaxy identifications are new, and one-quarter of the spectra show a 
single emission line with a large rest equivalent width ($\sim$100 \AA).  
}

\item{We report the discovery of F12$\_$ELG1 at $z$=0.7895, whose 
spectrum shows several Balmer emission lines indicative of a starburst 
phase for this young, low-mass galaxy with $\sim$solar metallicity and 
M/L$\approx$ 1.  Offset from the galaxy nucleus by $\sim$1kpc we
measure metal line ratios that are consistent with the presence of an AGN,
a result that is not apparent in the integrated spectrum.  We interpret the 
presence of a harder ionization source outside the galaxy nucleus to be a 
result of shocks possibly induced by a recent galaxy interaction.  
 }
 
 \item{We compute magnification factors for some individual galaxies.  We 
construct a magnification profile for the cluster, and measure a cumulative 
benefit due to lensing of a factor of 3.75 within the central $\approx$500 kpc.}

 \item
{We have detected the highest redshift galaxy with spectroscopic 
confirmation, ELG 10399 at $z$=5.13, and we report the discovery of 
several other objects, including F12$\_$ELG2 with mutiple sources of emission 
at close angular separations.}

\end{itemize}

Given their high magnifications over large areas, grism surveys in the fields
of massive lensing clusters are rich yet relatively unexplored territories.  In
particular lensing can open up the discovery space for galaxies with weak 
$W_r$ emission line sources.
For the typical  galaxy in the
background of the cluster, lensing causes both
the galaxy's
extended stellar continuum to be diluted and the star 
forming regions to be magnified but remain unresolved.  This lensing effect allows the 
detection threshold to be lowered
to include weaker $W_r$ ELSs.
The exquisite spatial resolution of HST
further allows for multiple weak ELSs to be identified in each ELG.
Operationally, such programs are free of lengthy selection algorithms, and so
can yield rewards even with only modest investments of telescope time.  

\acknowledgments
ACS was developed
Êunder NASA contract NAS 5-32865, and this research has been supported
in part
Êby NASA grant NAG5-7697 and by an equipment grant from Sun
ÊMicrosystems, Inc.
B. L. F. acknowledges generous hospitality at the SUNY Stony Brook Astronomy 
Group.
D. V. B. is funded through NASA Long-Term Space Astrophysics grant
NNG05GE26G.  We are grateful to Sangeeta Malhotra and James Rhoads for
providing useful follow-up spectroscopy on Magellan Observatories.  We
thank Holland Ford, Rogier Windhorst, Nicole Nesvadba, Jean-Paul Kneib
and Maru\u{s}a Brada\u{c} for useful discussions.  Some data for this
work was acquired at Keck Observatories.  We also want to express our appreciation
to the anonymous referee whose comments and suggestions improved 
the clarity of this paper.   The authors wish to extend
special thanks to those of Hawaiian ancestry on whose sacred mountain
we are privileged to be guests.  Without their generous hospitality,
some of the observations presented herein would not have been
possible.
 
\appendix
\section{The Catalog}

\begin{figure}
\includegraphics[viewport=-240 0 200 365,scale=0.5]{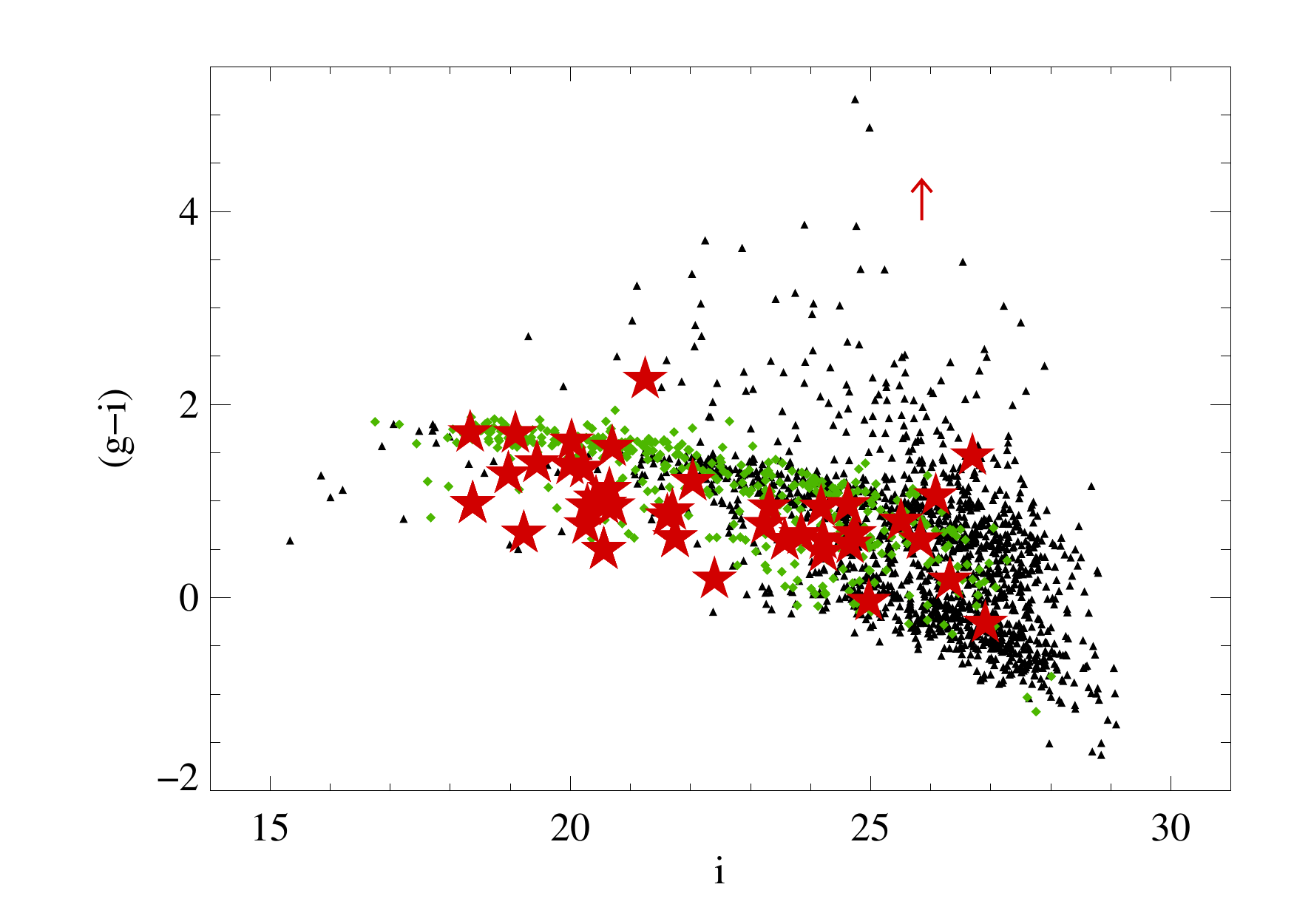}
\caption{Color-magnitude diagram for HST ACS $g_{475}$ and $i_{775}$ in AB
magnitudes.  Objects in our ELG sample are indicated by the red star-shaped 
symbols and compared with other objects, including photometric
redshifts for cluster members (green diamond shaped symbols) and 
photometric
for all other objects (black triangular shaped symbols).   The typical ELG in our
sample is bluer than the cluster.  The red arrow shows the lower limit in $(g-i)$ 
for the $g$-band dropout galaxy at $z$=5.13 ELG 10399.  \label{CMD}}
\end{figure}

In each emission line galaxy (ELG) we detect at least one 
 emission line source (ELS) with one or more emission lines (ELs).     
In all we identify 66 ELs in 52 ELSs in 43 ELGs.  
A color-magnitude diagram of our sample is shown in Figure~\ref{CMD}
(red star-shaped symbols), with the green diamond-shaped symbols
indicating the positions of cluster members obtained by photometric
redshifts.  The black triangle-shaped symbols show all objects in
the field with measured photometric redshifts.  Most objects in our sample have colors bluer than the
cluster sequence, corresponding to roughly $g-i=1.7$ at $i_{775} =
20$.  All nine H$\alpha$ Emitters (HAEs) in our sample that are
cluster members are also situated below the cluster sequence.  Four
ELGs are situated exactly on the cluster sequence.  These are not
cluster members and are identified as follows: ELG 11186 (HAE at
$z=0.235$), ELG 10746 (HAE at $z=0.230$), ELG 10226 (HAE at
$z=0.140$), and ELG 4277 (HAE at $z=0.462$).  Of these, ELG 11186 and
ELG 10746 are very close to the cluster, and may be foreground
galaxies experiencing infall into the cluster gravitational potential well.  The
three ELGs above the cluster sequence are: ELG 10399 at $z=5.12$ with
a lower limit of $g-i =4$, the disk galaxy ELG 1077 at $z=0.595$ with
$g-i=2.24$, and the faint and compact object ELG 6621 at $z=0.800$
with $g-i=1.48$.

 The spectroscopic results appear in Table~\ref{ID}
with the following columns: object name, right ascension and
declination, $i_{775}$ (AB) magnitude, emission line central
wavelength, emission line flux, total rest-frame equivalent width and
1-$\sigma$ errors, line identification, spectroscopic redshift, and a
reference to any redshifts in the literature determined
spectroscopically.  Eight line species are identified in our sample:
[S II]$\lambda\lambda$6716,6731, H$\alpha$, [O
  III]$\lambda\lambda$4959,5007, H$\beta$, [Ne III]$\lambda$3867, He
I$\lambda$4472, [O II]$\lambda\lambda$3727,3729, and Ly$\alpha$.  The
three doublets in this line list are unresolved at our spectral
resolution, so we adopt the following vacuum rest wavelengths for
their flux-weighted centroids: $\lambda_0$([S II]) = 6723.5 \AA,
$\lambda_0$([O III])=4996.5 \AA, and $\lambda_0$([O II])=3728.7 \AA.
Two people independently measured the rest-frame equivalent widths ($W_r$) 
and found consistent results.  The uncertainties reported for
the line fluxes are 1-$\sigma$ errors including continuum placement
and photon noise.  More than one-fourth of our ELGs have large rest
equivalent widths, $W> 100$ \AA.

\subsection{Line Identifications}
Previously published redshifts are the first resource for line identifications.
For new ELGs, the objects with multiple emission lines make the redshift
determination straightforward, taking into account the cases of similar
ratios of wavelengths between line pairs, 
such as the similar line ratios of $\lambda{H\alpha}/\lambda{[OIII]} $
and $\lambda{H\beta}/\lambda{[OII]}$ (M07).  The remainder 
of the catalog consists of single emission lines.  Single ELs present 
challenges as grism observations are not sensitive to small-scale changes 
that can serve as redshift-confirming features, such as strong absorption 
bands and the line shape of emission lines closely-separated in velocity space.  
The treatment of single emission lines relies on the
combination of photometric redshift, profile fitting and other sanity checks 
such as the search for continuum depressions and the absence of conflicting
emission features.  The photometric redshift is derived from a Bayesian 
approach, described in detail in other papers 
\citep{Benitez:00, Benitez:04, Coe:06}; it is measured including nine
bands and found to be a robust redshift indicator (see Fig.~9 in Frye et al.~2007).
We fit profiles to 
distinguish secondary bumps in our single emission lines to corroborate the 
likelihood of, for example, a bonafide [OIII]/H$\beta$ detection.  
When one emission line is observed, the photometric redshift is used to 
resolve the degenerate possible redshift solutions, resulting in a much
more accurate redshift than from the photometry alone.  

\begin{figure}
\includegraphics[viewport=-220 0 160 440, scale=0.52]{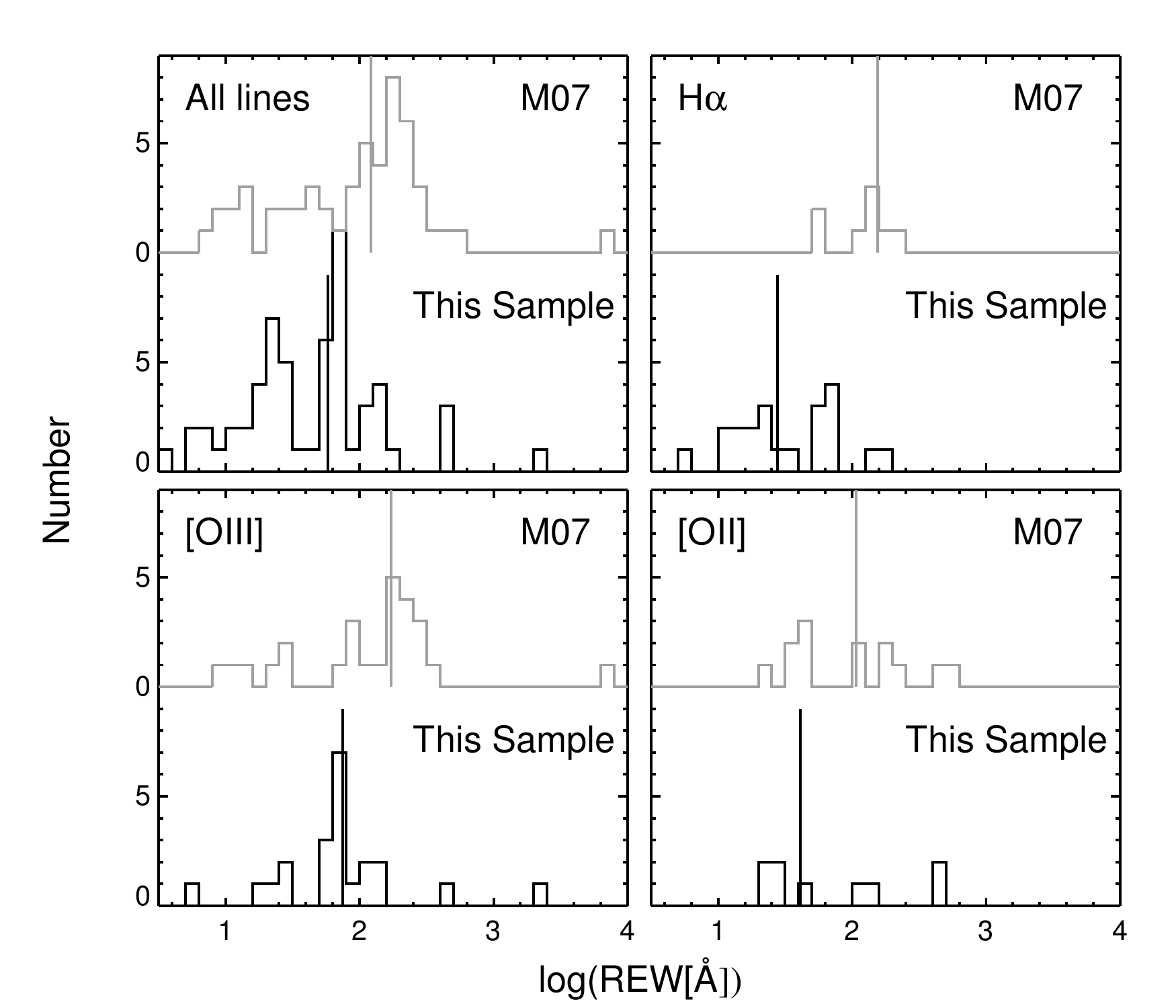}
\caption{The total rest-frame equivalent width ($W_r$) in Angstroms 
of each emission line in our catalog is plotted as a histogram for all objects, 
and then also for the three most common line species of H$\alpha$, [OII], and [OIII].
The results for this sample appear in black in the lower row of each panel and the
grism sample of the HDF-N (M07) in gray in the upper row. 
Our lensed sample includes more weak $W_r$ sources of line emission. 
The effect of lensing is to boost the flux of the unresolved emission
lines with respect to the stretched and diluted stellar continuum,
thereby increasing the signal-to-noise and allowing for lower
$W_r$ detection thresholds.   \label{figREW}}
\end{figure}

The largest source of uncertainty is in establishing the goodness of
fit of the zero-point of the wavelength solution.  In our blind
emission line finding technique, ELs are identified and the offset of
the position between the emission line and the direct source are
computed.  The goodness of fit can be measured by comparing the grism
and spectroscopic redshifts.  From previous studies involving similar
datasets and reduction techniques as described in M07, the scatter in
grism and spectroscopic redshifts is equivalent to an error in
position of $\sim$1 pixel, or $\Delta v$=1650 km s$^{-1}$.  In
addition, the spectral resolution is low, with $R$=90 at
$\lambda$=8500 \AA, corresponding to $\Delta v$=3300 km s$^{-1}$ or
about 2.1 pixels.  While in practice this means that we will have
relatively little leverage on measuring kinematics of gas clouds
within a disk, the grism has the advantage of obtaining spectroscopy
that is {\it spatially} resolved on impressively small angular scales
of $\sim0\farcs2$

\subsection{Line Demographics}

 We compare line properties sorted by the three most common line species,
 H$\alpha$, [OIII], [OII] (Table~\ref{ID}), and compare these results 
with a recent HST ACS grism survey
in an unlensed field to similar depth, the HDF grism survey (M07).
By line species, H$\alpha$ emitters are plentiful despite the small
sample volume owing to the field selection of a massive galaxy
cluster.  The median redshift for our sample of HAEs
of $z=0.22$ is close to the cluster redshift of $z=0.187$ and is
dominated by galaxies in the cluster.
We measure the median $W_r$s to be: $W_{H\alpha}$ = 29.3 \AA,
$W_{[OIII]}$ = 67.2 \AA, and $W_{[OII]}$ = 29.6 \AA.  These values are
large, but typical of grism surveys.  

The largest difference between
this sample and the HDF is in the values of the $W_r$.  One quarter of
all of the emitters in our sample have $W_r$$\ge$100 \AA, compared to
three-quarters for the comparison sample  (HDF-N, M07).  Also, the median 
$W_r$ for each line species is smaller than the values in the comparison 
sample (Fig.~\ref{figREW}).   The largest shift is found for the HAEs, which 
are dominated by galaxies in or near to this old and presumably relaxed 
cluster ($z$=0.19).   Interestingly, all of the HAEs in our sample that are 
brighter than the magnitude limit of the large emission line survey of B02 
of $I$=19.3 are recovered in our census.  Moreover, all of the HAEs in our 
sample that are fainter than their limiting magnitude are new to the literature.  
Thus we find the slitless grism observing approach to be especially sensitive 
to the detection of low-luminosity (and largely unlensed) HAEs with low SFRs.  
On the other hand, the shift in the [OIII] and [OII] emitters towards lower $W_r$s 
is owing to lensing.  Lensing boosts the brightnesses and sizes of the background
objects, including the emission lines, with the effect of improving the signal to 
noise of the grism spectra.  This effect enables the identification of weaker 
emission lines with smaller $W_r$ and lower SFR than are found in the field. We
compute rest-frame $B$-band magnitudes, $M_B$, by fitting a Bruzual-Charlot 
model to our optical ACS photometry, and then computing the k-correction onto 
a Johnson $B$-band filter template.  The mean $M_B$ for the 27 [OIII] emitters 
is fainter than for the HDF by 1.5 magnitudes, yielding yet another indication of 
the gravitational lensing effect.  The shift is not seen for the 21 H$\alpha$ emitters 
or nine [OII] emitters which is dominated by emitters inside the cluster lens and
small number statistics, respectively.

\subsection{Special Cases}
 

 Seven emission lines (ELs) flagged by our purpose-built reduction
 code were later removed as bogus detections.  In five cases the lines
 were weak and other available spectroscopic data ruled out the grism
 features as ELs.  In two cases, the continuum shapes mimicked an EL
 but turned out to be stellar continuum.
There are three cases for which the EL redshifts are not consistent
with published redshifts: ELS 1184, ELS 11324, and ELG 6621.  We discuss
these cases below.

\begin{itemize}

\item{
ELS 1184 is in a crowded environment and is separated by only 
1.8$\arcsec$ from ELS 1157 (see upper left of Fig.~\ref{fighiz}).  
 ELS 1157 has multiple
emission features confirming the redshift of $z=0.195$ for this
previously identified object \citep{Balogh:02, Duc:02}.  ELS 1184 has
a single EL at 6917 \AA \ that we do not identify with any strong
emission feature at or near to $z\sim0.195$.  We adopt the foreground
redshift $z=0.054$ to this compact and very bright ELS based on the
absence of other expected ELs in the bandpass were the identifications
to correspond to [OIII] or [OII].  We measure a photometric redshift
for the Sc galaxy of $z_{BPZ} = 0.320 \pm 0.129$, but this measurement
is taken at the position of ELS 1157.}
 
\item{
ELG 11324 is in the crowded outer regions of an extended face-on
spiral galaxy with a photometric redshift of $z_{BPZ} =
0.250^{+0.260}_{0.122}$ and a spectroscopic redshift of $z=0.384$
(J. Richard, {\it private communication}).  We detect a single EL at
6913.5 \AA \ which we take to be [OIII] at the spectroscopic redshift.}

\item{ELG 6621 is small and faint, with a magnified size of
$0\farcs$3 and $i_{775} = 26.7$. The grism spectrum shows a lone
emission line at 6712.4 \AA \ with stellar continuum and no salient
continuum features.  This galaxy suffers from crowding by a second
object that is confidently identified from our companion Magellan
spectrum as ELG 6621a at $z=0.1868$.  ELG 6621a is a cluster
member that is not detected in our grism dataset and that has no feature at the
position of the grism EL.  The photometric redshift for ELG 6621 is $z_{BPZ}$ =
$3.780 \pm 0.468$.  If the EL in ELG
6621 is Ly$\alpha$ then the redshift is $z$=4.52, and we cannot rule
out the possibility of high redshift at least according the redshift
test available to us \citep{Rhoads:09}.  The high redshift scenario
would seem unlikely given that the photometry does not correct for 
 extinction at the cluster redshift, and thus we must attribute some of its `redness' to its 
 being behind the foreground cluster member.  
Strictly, the EL can be also H$\alpha$, [OII] or [OIII].  
 If the line is [OIII] then $z$=0.346, in which case we do not detect
 the non-requisite H$\beta$ EL.  Based on the large observed equivalent
 width of 772.9 $\pm$ 497.9 we adopt a line identification of [OII] at $z=0.800$, and admit this as an uncertain redshift in our sample.  Note ELG 6621a is detected
 only in our Magellan spectrum and so does not appear in Table 1.}

\end{itemize}


\begin{deluxetable}{ c c c c c c c c c c}
\tabletypesize{\scriptsize}
\tablecaption{Spectroscopic Catalog of Galaxies in the Field of A1689}
\tablewidth{0pt}
\tablehead{
\colhead{ID} & \colhead{RA} & \colhead{DEC}  
& \colhead{$i_{775}$} & \colhead{Line Center} & \colhead{log$(f_{l})$} & \colhead{REW} & \colhead{Line ID} & \colhead{$z_{grism}$}  & \colhead{Ref} \\
\colhead{ }&\colhead{(hours)}&\colhead{($\circ$)}&\colhead{(AB)} &\colhead{(\AA)} & \colhead{($erg \ s^{-1} \ cm^{-2}$) } 
& \colhead{(\AA)}&\colhead{ }&\colhead{ } & \colhead{} \\
}

\startdata


 
6621 & 13.189628 & -1.3506333   
 & $26.69 \pm 0.04$& 6712.4 & -16.63 
&$429 \pm 277$ 
& [OII] & 0.800 & \\ 

3483 & 13.189729   &   -1.323381 
& $20.11 \pm 0.01$ & 8921.4 &  -15.25 
& $-$
& HeI & 0.04 & \\ 

11260 & 13.190046   &   -1.348567 
& $24.21 \pm 0.01$& 9010.5 & -15.74 
&$ 2034.8 \pm 483.0$
& [OIII] & 0.810 & \\ 

 6381 & 13.190094   &   -1.341133 
 & 24.17 $\pm$ 0.01& 8740.6 &  -16.06 
&$ 97.2 \pm 19.5$
& [OIII] & 0.758 & \\ 

11186 & 13.190131 &    -1.353053 
& 18.33 $\pm$ 0.01 & 8126.7 & -15.86 
&$6.1 \pm 1.4$ 
& H$\alpha$ & 0.235 & a \\ 

11136 & 13.190185  &   -1.357363 
& $23.32 \pm 0.01$ & 8045.1 & -16.35 
&$65.8 \pm 15.2$ 
& [OIII] & 0.615 & \\ 

11322 &  13.190542  &    -1.326375 
& $19.82 \pm 0.01$& 9505.6 & -16.18 
&$ 67.5 \pm 16.1$
& H$\alpha$ & 0.449 & a \\ 

6182 & 13.190737   &   -1.324731 
&$ 24.66 \pm 0.01$& 8383.4 & -16.54 
&$ 59.4 \pm 20.0$
&H$\alpha$ & 0.277 & \\ 

11226 & 13.190740   &   -1.3298028 
&  $21.61 \pm 0.01$ & 7783.1 & -16.61 
&$13.3 \pm 5.0$
& H$\alpha$ & 0.186 & \\ %


11040 & 13.190872   &   -1.349414 
& $18.97 \pm 0.01 $ & 7983.6 & -16.09 
&$10.7 \pm 2.7$ 
& H$\alpha$ & 0.215 & a, b \\ 

4752 &  13.190875 &     -1.3495417 
 & $18.97 \pm 0.01 $ & 7956.9 & -16.06 
& $10.8 \pm 1.3$ 
& H$\alpha$ & 0.212 & a, b \\ 

6680	 & 13.190824   &   -1.311189 
& $ 25.83 \pm 0.03$ & 7332.8 & -16.59 
&$<87$  & [OIII] & 0.470 & \\ 

5582	 & 13.190834  &    -1.334439 
& 26.32  $\pm$  0.04 &  7531.3 & -16.42 
&$66.3 \pm 17.9$
& [OIII] & 0.510 & \\ 

11324 & 13.190896   &   -1.313931 
& $21.70 \pm 0.01$ & 6913.5 & -16.55 
&$54.6 \pm 15.1$
& [OIII] & 0.384 & c \\ 

4971	 &  13.190906   &   -1.349731 
& $18.97 \pm 0.01 $ & 7972.9 & -15.89 
&$64.1 \pm 8.7$ 
& H$\alpha$ & 0.202 & a, b \\  

6583	 & 13.190941   &   -1.309417 
&23.84 $\pm$ 0.01 & 7932.0 &  -16.69 
&$<50$  & [OIII] & 0.590 & \\ 

 6578 & 13.190947   &   -1.309414 
 &  23.84  $\pm$  0.01& 7917.2 &  -16.19 
&$<137$   & [OIII] & 0.590 & \\ 


 4277 & 13.191041    &  -1.352267 
  & $20.70 \pm 0.01$ & 9571.7 &  -15.62 
&$18.3 \pm 1.7$ 
& H$\alpha$ & 0.462  & \\ 
   
 4298 & 13.191076 &     -1.351167 
 & 21.74$\pm$  0.01 & 7933.4 & -15.74 
&$70.4 \pm 5.5$ 
& H$\alpha$ & 0.209 & \\  

$^{\prime\prime}$&  $^{\prime\prime}$  & $^{\prime\prime}$  
& $^{\prime\prime}$ & 6012.6 & -15.77 
&$17.5 \pm 2.2$ 
& [OIII] & $^{\prime\prime}$  & \\ 

 $^{\prime\prime}$  & $^{\prime\prime}$  & $^{\prime\prime}$ 
 & $^{\prime\prime}$ & 8128.4 &  -16.20 
&$18.1 \pm 2.6$ 
& [SII] & $^{\prime\prime}$  & \\ 

4251 &  13.191098   &   -1.350869 
  & $21.74 \pm 0.01$ &	7978.2 & -16.21 
&$57.8 \pm 12.1$ 
& H$\alpha$ & 0.210 & \\ 
 
  5700 &  13.191123  &    -1.321883 
 & $22.04 \pm 0.01$& 7966.7 & -15.57 
&$107.3 \pm 6.0$
& [OII] & 1.139 & c \\ 
   
$^{\prime\prime}$  & $^{\prime\prime}$  & $^{\prime\prime}$  
& $^{\prime\prime}$  & 8235.6 & -16.41 
&$ 24.3 \pm 3.6$
& [NeIII] & $^{\prime\prime}$  & \\ 

5570	 & 13.191162   &   -1.323636 
&  23.57 $\pm$   0.01& 8155.7 & -16.19 
&$73.4 \pm 16.4 $
& [OIII] & 0.637 & \\ 

 11149 &  13.191176   &   -1.323664 
& 23.57 $\pm$   0.01 & 8115.4 & -16.23 
&$144.2 \pm 42.4$ 
& [OIII] & 0.629 & \\ 

6154	 &  13.191220    &  -1.309083 
&  23.24 $\pm$ 0.01 & 7344.0 & -16.31 
&$127.2 \pm 21.9$
& [OIII] & 0.473 & a \\ 

 4744 &  13.191277    &  -1.338045 
 & $24.73 \pm 0.02$ & 8826.7 & -15.74 
&$149.3 \pm 11.6$ 
& [OII] & 1.368 & \\

10640\tablenotemark{a} &  13.191326  &  -1.361966 
&$20.56 \pm 0.01$ & 6656.1  &  -16.05 
&$41.2 \pm 6.6$ 
&[OII] & 0.783 & a, d \\  

$^{\prime\prime}$ & $^{\prime\prime}$  & $^{\prime\prime}$  
&$^{\prime\prime}$ & 8928.1 & -16.24
&$74.5 \pm 19.3$ 
&[OIII] & 0.787  & \\ 

10638\tablenotemark{a}  &13.191334    &  -1.362022 
& $^{\prime\prime}$& 6753.2 & -16.16 
&$26.1 \pm 5.0$ 
& [OII] & 0.820 & a, d \\ 

$^{\prime\prime}$ & $^{\prime\prime}$  & $^{\prime\prime}$  
& $^{\prime\prime}$ & 9020.8 & -16.27 
&$77.5 \pm 7.7$ 
&[OIII] & 0.807  &  \\ 

20002\tablenotemark{a}  &  13.191347    &  -1.362072 
& $^{\prime\prime}$ & 6673.2 & -16.31 
&$24.1 \pm 4.1$ 
& [OII] & 0.790 & a, d \\ 

 $^{\prime\prime}$ & $^{\prime\prime}$  & $^{\prime\prime}$ 
& $^{\prime\prime}$ & 8935.5 & -15.98 
&$77.8 \pm 9.8$ 
& [OIII] & $^{\prime\prime}$  &  \\ 

 2630	  &  13.191381  &    -1.363211 
& 24.97 $\pm$   0.02& 6658.2 & -16.44 
&$104.3 \pm 26.8$ 
& [OIII] & 0.335 & \\ 

2494  &  13.191546   &  -1.360030 
& $24.21 \pm 0.01$ & 6071.8 & -15.86 
&$407.9 \pm 66.3$ 
& [OIII] & 0.215 & \\ 
 
 $^{\prime\prime}$ & $^{\prime\prime}$   & $^{\prime\prime}$ 
& $^{\prime\prime}$  &7970.2 & -16.35 
&$176.3 \pm 45.9$ 
&H$\alpha$ & $^{\prime\prime}$ & \\ 

10746 &  13.191596   &   -1.350106 
& $19.08\pm 0.01$ & 8069.4 & -16.09 
& $21.2 \pm 3.2$ 
& H$\alpha$ & 0.230 & \\ 

11085 &  13.191659    &  -1.317806 
& $20.43  \pm 0.01 $ & 7652.2 & -15.68 
&$61.4 \pm 4.2$ 
& [OIII] & 0.532 & \\ 

20004 & 13.191663   &   -1.317805 
& $20.43  \pm 0.01$ & 7790.0 &  -16.15 
&$ 28.4 \pm 1.7$ 
&[OIII] & 0.560 & \\ 


$^{\prime\prime}$  & $^{\prime\prime}$  & $^{\prime\prime}$ 
& $20.43  \pm 0.01$   & 7560.1 & -15.86 
&$29.3 \pm 2.3$
&[OIII] & 0.513 &  \\ 

5158	 &  13.191787    &  -1.313802 
&  24.62$\pm$   0.01& 8332.8 & -15.67 
& $< 296$  
& [OIII] & 0.671 &    \\ 

$^{\prime\prime}$  & $^{\prime\prime}$  & $^{\prime\prime}$  
& $^{\prime\prime}$  & 8123.4 & -16.80 
&\nodata & H$\beta$ & $^{\prime\prime}$ & \\ 

1946 &  13.191900  &   -1.360658 
& $20.57\pm0.01$ & 8419.0 & -15.71 
&$78.5 \pm 12.1$ 
& [OIII] & 0.700 & a, e \\ 

$^{\prime\prime}$  & $^{\prime\prime}$   & $^{\prime\prime}$   
&$^{\prime\prime}$ & 6352.2 & -16.05 
&$20.7 \pm 4.7$ 
&[OII] & $^{\prime\prime}$ &   \\ 

 4194 & 13.191980   &   -1.323292 
 & $22.40 \pm 0.01$ & 8001.5 & -16.31 
&$29.1 \pm 7.1$ 
& [OII] & 1.145 & c \\ 

 10412 &  13.192066  &  -1.359247 
 & $20.65\pm0.01$ & 7736.1 & -15.87 
&$34.5 \pm 8.0$ 
& H$\alpha$ & 0.179 & \\ 
 
 10782 & 13.192252  &    -1.326700 
& $19.22 \pm0.01$ & 7566.0 & -15.88 
& $27.6 \pm 2.4$ 
& H$\alpha$ & 0.153 & b \\ 

3203	 &  13.192258   &   -1.326697 
& $19.22\pm0.01$ & 7678.8 &  -15.26 
&$72.4 \pm 3.3$ 
& H$\alpha$ & 0.170 & b \\ 

$^{\prime\prime}$ & $^{\prime\prime}$  & $^{\prime\prime}$   
& $^{\prime\prime}$ & 7866.8 &  -16.21 
&$7.1 \pm 1.0$ 
& [SII] & $^{\prime\prime}$ & \\ 

10154 & 13.192360 & -1.371497 
&20.26 $\pm$  0.01 & 6715.2 & -16.00 
&$59.5 \pm 9.2$
&[OIII] & 0.115 & a \\ 

10226 & 13.192504 & -1.361381 
&20.02  $\pm$ 0.01 & 7479.4 & -16.60 
& $19.7 \pm 7.1$
&H$\alpha$ & 0.140 & \\ 

1651 &  13.192538 &  -1.3473694 
&25.51$\pm$0.02 & 8850.5 &  -15.94 
&$434.2 \pm 114.0$ 
& [OII] & 1.38 & \\ 

804      & 13.193044 & -1.351555 
&26.91   $\pm$ 0.07 & 6588.0 & -16.56 
&$106.1 \pm 33.6$ 
&[OIII] & 0.320 &  \\ 

10399 & 13.193055 & -1.330869 
& $25.85\pm0.02$ & 7445.9 &-16.63
&$< 100$ & Ly$\alpha$ & 5.13& e, f \\ 

1077   & 13.193131 & -1.341633 
& 21.25  $\pm$ 0.01 & 7967.6 & -16.17 
&$23.2 \pm 2.8$
&[OIII] & 0.595 & a, e \\ 

 1184 & 13.193212 & -1.337078 
& 18.37$\pm$ 0.01  & 6917.1 &  -15.42 
&$61.3 \pm 5.5$
& H$\alpha$ & 0.054 & a, b \\ 

1157	 & 13.193232 & -1.336661 
&18.37$\pm$ 0.01& 7852.0 & -14.82 
&$133.6 \pm 6.0$ 
&H$\alpha$ & 0.195 & a, b \\ 


$^{\prime\prime}$ & $^{\prime\prime}$   & $^{\prime\prime}$ 
& $^{\prime\prime}$ & 5970.9 & -15.78 
&$5.5 \pm 1.1$ 
&[OIII] & $^{\prime\prime}$  &  \\ 

$^{\prime\prime}$ & $^{\prime\prime}$   & $^{\prime\prime}$ 
& $^{\prime\prime}$ & 5795.5 & -15.53 
&$3.5 \pm 1.3$
&H$\beta$& $^{\prime\prime}$ &  \\ 

$^{\prime\prime}$ & $^{\prime\prime}$   & $^{\prime\prime}$ 
& $^{\prime\prime}$ & 7040.1 & -15.94 
&$8.6 \pm 2.1$
& HeI& $^{\prime\prime}$ &  \\ 

1507	 & 13.193252 & -1.326861 
& $20.22\pm0.01$ & 7833.0 & -16.37 
& $15.7 \pm 3.3$
& H$\alpha$ & 0.194 & \\ 


1094  & 13.193518 & -1.328469 
& $19.44\pm0.01$  & 7848.0 & -15.89 
& $23.7 \pm 2.2$
& H$\alpha$ & 0.195& b \\ 

$^{\prime\prime}$  & $^{\prime\prime}$ & $^{\prime\prime}$
& $^{\prime\prime}$& 8035.6 & -16.38 
&$6.7 \pm 1.2$ 
&[SII] & $^{\prime\prime}$ & \\ 

486  	   & 13.193702 & -1.335958 
& 26.08 $\pm$   0.03 & 8057.7 & -16.57 
&$< 92$ & [OIII] & 0.615 &  \\ 

20001 & 13.193894 & -1.335958 
&20.28$\pm$  0.01& 7901.7 & -16.11 
&$20.2 \pm 2.2$ 
&H$\alpha$ & 0.205 & b \\ 

\enddata%
\tablenotetext{a}{Spectroscopic redshift from the catalog of \citet{Duc:02}}%
\tablenotetext{b}{Spectroscopic redshift from the catalog of \citet{Balogh:02}}%
\tablenotetext{c}{Spectroscopic redshift from Richard et al. 2008, private communication.}
\tablenotetext{d}{This is an emission line in the ELG F12$\_$ELG1.}
\tablenotetext{e}{Spectroscopic redshift from the of \citet{Frye:07}}
\tablenotetext{f}{Spectroscopic redshift from the catalog of \citet{Frye:02}}
\label{ID}
\end{deluxetable}

\clearpage

\begin{deluxetable}{c c c}
\tabletypesize{\scriptsize}
\tablecaption{ELGs with Multiple Emission Line Sources}
\tablewidth{0pt}
\tablehead{
\colhead{Group} & \colhead{ID} & \colhead{$z_{ELS}$} \\
}

\startdata
   1\tablenotemark{a} & 10782 & 0.153 \\
       & 3203 &  0.170 \\
  2 (F12$\_$ELG2) & 20004 & 0.560 \\
     &  11085 & 0.532  \\
 3 & 4971    & 0.202   \\
     & 4752   &  0.212  \\
     & 11040 & 0.215   \\
  4\tablenotemark{b}   & 4298  & 0.209  \\
     & 4251  & 0.210   \\
5 & 6583   & 0.590    \\
    & 6578   & 0.590   \\
6    & 11149 & 0.629  \\
  & 5570   & 0.637      \\
7 (F12$\_$ELG1)   & 10638 ``B"  & 0.815  \\  
    & 20002 ``A"   & 0.790  \\ 
     & 10640 ``C"  & 0.785  \\
\enddata\
\tablenotetext{a}{ELS 10782 is in the foreground of A1689.}
\tablenotetext{b}{There is a third ELS projected near to
the line of sight but with different colors, ELS 4277 at $z_{grism}$ = 0.462.} 
\label{multiples}
\end{deluxetable}

\clearpage

\begin{deluxetable}{ c c}
\tabletypesize{\scriptsize}
\tablecaption{Intrinsic Fluxes for F12$\_$ELG1}
\tablewidth{0pt}
\tablehead{
\colhead{Line} & \colhead{Line Flux\tablenotemark{a}} \\
\colhead{}        & \colhead{(ergs s$^{-1}$ cm$^{-2}$)} \\
}

\startdata
[O II]                              & $1.2^{+0.34}_{-0.27} \times 10^{-15}$  \\  

[O III]\tablenotemark{b} & $1.8^{+0.17}_{-0.32} \times 10^{-15}$ \\ 

[Ne III]                            & $6.0^{+0.12}_{-4.2} \times 10^{-17}$  \\ 

H$\beta$\tablenotemark{c}                        & $3.0^{+1.1}_{-1.2} \times 10^{-16}$    \\ 

H$\gamma$\tablenotemark{c}                   & $1.0^{+1.0}_{-0.88} \times 10^{-16}$   \\
\enddata
\tablenotetext{a}{Line fluxes are measured from the Keck spectroscopy and are corrected for extinction and cluster magnification.}

\tablenotetext{b}{[O III]$\lambda$4959 is contaminated by night sky lines, so we adopt
the line ratio [O III]$\lambda\lambda$4959, 5007 = $1.3 \times$ [O III]$\lambda$5007.  The flux value is taken from the Gaussian fit to the line.}
\tablenotetext{c}{Balmer line fluxes are additionally corrected for underlying stellar absorption.}
\label{fluxes}
\end{deluxetable}

\clearpage

\begin{deluxetable}{c c c c c c c c}
\tabletypesize{\scriptsize}
\tablecaption{Objects with Spectroscopy at $z>2.5$ in the A1689 Field}
\tablewidth{0pt}
\tablehead{
\colhead{RA} & \colhead{DEC} & \colhead{Redshift} & \colhead{Image Size\tablenotemark{a}} &\colhead{Intrinsic Size\tablenotemark{b}}  & \colhead{$i_{775}$} & \colhead{$W_{Ly\alpha}$} & Ref \\
\colhead{(hours)}&\colhead{($\circ$)}& \colhead{} & \colhead{($^{\prime \prime})$}  & \colhead{($h^{-1}$kpc )} &\colhead{ $(AB)$} &\colhead{(\AA)} & \colhead{} \\
}
\startdata
13.190701 & -1.3320694 & 3.038 & 1.3 & 0.08 & 23.07 $\pm$ 0.01 & 4.0$^{+1.5}_{-5.0}$ & c, d \\ 
13.190681 & -1.3324306 & 3.038 & 2.2  &  1.1     & 22.40 $\pm$ 0.01 & -26$^{-5.4}_{+1.7}$           & c, d \\ 
13.192517 & -1.3409444 & 3.038 & 0.75 &  0.73     & 24.02 $\pm$ 0.01 & -4.0$^{-1.5}_{+11.5}$   & c, d \\ 
13.191649  &  -1.3207361 & 3.770 &  2.3   & 0.22        & 24.13 $\pm$ 0.90 & $\leq3.3 $                      & d, e \\ 
13.190402 & -1.3477333  & 4.868 &  1.3  & 0.13         & 23.48 $\pm$ 0.03 &12.4$^{+8.83}_{-3.84}$ & d, e, f \\  
13.193053 & -1.3308528 & 5.120  &  0.64   & $\sim1$  & 25.58 $\pm$ 0.18 & 29.7$^{+13.29}_{-4.6}$& d, e, g \\ 
\enddata
\tablenotetext{a}{Image size is measured from the HST ACS $i_{775}$ image directly, without
applying a lensing correction factor.}
\tablenotetext{b}{Linear size is computed from the image size, given a lensing magnification factor and our adopted cosmology.}
\tablenotetext{c}{\citet{Broadhurst:05}}
\tablenotetext{d}{\citet{Frye:07}}
\tablenotetext{e}{\citet{Frye:02}}
\tablenotetext{f}{\citet{Frye:08}} 
\tablenotetext{g}{This paper.}
\label{tablehiz}
\end{deluxetable}

\clearpage

\begin{deluxetable}{c c c c}
\tabletypesize{\scriptsize}
\tablecaption{Properties of ELSs by Line Species}
\tablewidth{0pt}
\tablehead{
\colhead{} & \colhead{H$\alpha$} & \colhead{[OIII]} & \colhead{[OII]} \\
}
\startdata
Number  ELSs                &  21\tablenotemark{a}       & 27\tablenotemark{b}     & 9\tablenotemark{c} \\
Mean $z$                & 0.22   & 0.51    & 0.92 \\ 
Mean $M_B$          & -18.1    & -18.6  & -20.8 \\ 
Mean $W_r$\tablenotemark{d} (\AA)   & 40.4 & 170 & 137  \\
\enddata
\tablenotetext{a}{There are 18 ELGs that are H$\alpha$ emitters.}
\tablenotetext{b}{There are 22 ELGs that are [OIII] emitters.}
\tablenotetext{c}{There are 7 ELGs that are [OII] emitters.}
\tablenotetext{d}{Given large uncertainties in the background subtraction, we report 
$W_r$s for only 23 [OIII] emitters.}
\label{stats}
\end{deluxetable}

\end{document}